\documentclass[12pt]{article}
\usepackage{graphicx}

\def\hybrid{\topmargin 0pt      \oddsidemargin 0pt
        \headheight 0pt \headsep 0pt
       \voffset-1cm
        \textwidth 6.25in       
       \textheight 9.5in       
        \marginparwidth 0.0in
        \parskip 5pt plus 1pt   \jot = 1.5ex}
\catcode`\@=11
\def\marginnote#1{}

\newcount\hour
\newcount\minute
\newtoks\amorpm
\hour=\time\divide\hour by60
\minute=\time{\multiply\hour by60 \global\advance\minute by-\hour}
\edef\standardtime{{\ifnum\hour<12 \global\amorpm={am}%
        \else\global\amorpm={pm}\advance\hour by-12 \fi
        \ifnum\hour=0 \hour=12 \fi
        \number\hour:\ifnum\minute<10 0\fi\number\minute\the\amorpm}}
\edef\militarytime{\number\hour:\ifnum\minute<10 0\fi\number\minute}

\def\draftlabel#1{{\@bsphack\if@filesw {\let\thepage\relax
   \xdef\@gtempa{\write\@auxout{\string
      \newlabel{#1}{{\@currentlabel}{\thepage}}}}}\@gtempa
   \if@nobreak \ifvmode\nobreak\fi\fi\fi\@esphack}
        \gdef\@eqnlabel{#1}}
\def\@eqnlabel{}
\def\@vacuum{}
\def\draftmarginnote#1{\marginpar{\raggedright\scriptsize\tt#1}}

\def\draftlabel#1{{\@bsphack\if@filesw {\let\thepage\relax
   \xdef\@gtempa{\write\@auxout{\string
      \newlabel{#1}{{\@currentlabel}{\thepage}}}}}\@gtempa
   \if@nobreak \ifvmode\nobreak\fi\fi\fi\@esphack}
        \gdef\@eqnlabel{#1}}
\def\@eqnlabel{}
\def\@vacuum{}
\def\draftmarginnote#1{\marginpar{\raggedright\scriptsize\tt#1}}

\def\draft{\oddsidemargin -.5truein
        \def\@oddfoot{\sl preliminary draft \hfil
        \rm\thepage\hfil\sl\today\quad\militarytime}
        \let\@evenfoot\@oddfoot \overfullrule 3pt
        \let\label=\draftlabel
        \let\marginnote=\draftmarginnote
   \def\@eqnnum{(\theequation)\rlap{\kern\marginparsep\tt\@eqnlabel}%
\global\let\@eqnlabel\@vacuum}  }

\def\numberbysection{\@addtoreset{equation}{section}
        \def\theequation{\thesection.\arabic{equation}}}

\def\underline#1{\relax\ifmmode\@@underline#1\else
        $\@@underline{\hbox{#1}}$\relax\fi}

\def\titlepage{\@restonecolfalse\if@twocolumn\@restonecoltrue\onecolumn
     \else \newpage \fi \thispagestyle{empty}\c@page\z@
        \def\thefootnote{\fnsymbol{footnote}} }

\def\endtitlepage{\if@restonecol\twocolumn \else  \fi
        \def\thefootnote{\arabic{footnote}}
        \setcounter{footnote}{0}}  
\relax

\numberbysection
\hybrid

\newfont{\Bbb}{msbm10 scaled 1\@ptsize00}
\newfont{\Bbbb}{msbm7 scaled 1\@ptsize00}

\newcommand{\DDD}{\raise-1pt\hbox{$\mbox{\Bbbb D}$}}

\newcommand{\UUU}{\raise-1pt\hbox{$\mbox{\Bbbb U}$}}

\newcommand{\z}{\raise-1pt\hbox{$\mbox{\Bbbb Z}$}}

\newcommand{\sss}{\raise-1pt\hbox{$\mbox{\Bbbb S}$}}

\def\beq{\begin{equation}}
\def\eeq{\end{equation}}
\def\p{\partial}

\newtheorem{lemma-definition}{Lemma-Definition}[section]

\newtheorem{proposition}{Proposition}[section]

\def\res{\mathop{\hbox{res}}\limits}

\def\square{\hfill
{\vrule height6pt width6pt depth1pt} \break \vspace{.01cm}}

\begin{document}

\begin{titlepage}

\title{Monodromy free linear equations 
and many-body systems}

\author{I. Krichever\thanks{
Columbia University, New York, USA;
e-mail: krichev@math.columbia.edu}
\and
A.~Zabrodin\thanks{
Skolkovo Institute of Science and Technology, 143026, Moscow, Russia and
Steklov Mathematical Institute of Russian Academy of Sciences,
Gubkina str. 8, Moscow, 119991, Russian Federation;
e-mail: zabrodin@itep.ru}}

\date{November 2022}
\maketitle


\begin{abstract}

We further develop the approach to many-body systems based on finding
conditions of existence of meromorphic solutions to certain linear
partial differential and difference
equations which serve as auxiliary linear problems for nonlinear
integrable equations such as KP, BKP, CKP and different versions
of the Toda lattice. These conditions imply 
equations of the time evolution for poles of singular
solutions to the nonlinear equations which are equations of motion for
integrable many-body systems of Calogero-Moser and Ruijsenaars-Schneider 
type. A new 
many-body system is introduced, which governs dynamics of poles 
of elliptic solutions to the Toda lattice of type B.

\end{abstract}

\end{titlepage}

\vspace{5mm}

%

\tableofcontents

\vspace{5mm}

\section{Introduction}

Dynamics of poles of singular solutions to nonlinear integrable equations
is a well-known subject in the theory of integrable systems. Investigations
in this direction were initiated in the seminal paper \cite{AMM77}.
In \cite{Krichever78,CC77} it was shown that poles $x_i$ 
of rational solutions
to the Kadomtsev-Petviashvili (KP) equation move as particles of the
many-body Calogero-Moser system \cite{Calogero71,Calogero75,Moser75} 
with the pairwise potential
$1/(x_i-x_j)^2$.
This remarkable connection
was further generalized to elliptic (double periodic) 
solutions in \cite{Krichever80}:
poles $x_i$ of the elliptic solutions were shown to 
move according to the equations of motion
of Calogero-Moser particles with the elliptic
interaction potential $\wp (x_i-x_j)$, where 
$\wp$ is the elliptic Weierstrass $\wp$-function.
This many-body system of classical mechanics is known to be integrable. 
For a review of the models of the Calogero-Moser type see \cite{OP81}.

The correspondence between 
singular solutions of nonlinear integrable 
equations and integrable many-body systems 
allows for generalizations in various directions. 
The extension to the matrix KP equation was discussed in 
\cite{KBBT95}; in this case 
the poles and matrix residues at the poles move as particles of the spin generalization
of the Calogero-Moser model known also as the 
Gibbons-Hermsen model \cite{GH84}. In the paper 
\cite{KZ95} by the authors, the dynamics of poles of elliptic
solutions to 
the (matrix) two-dimensional
Toda lattice was shown to be isomorphic to the  
relativistic 
extension of the Calogero-Moser system
known also as the Ruijsenaars-Schneider system \cite{RS86,Ruij87}
and its version with spin degrees of freedom \cite{KZ95}. 
Another generalizations concern 
B- and C-versions of the KP equation (BKP and CKP). Equations of motion
for poles of elliptic solutions to the BKP and CKP equations were 
recently obtained in \cite{RZ20} and \cite{KZ20} respectively.
For a review see \cite{Z19}. 

The method suggested in \cite{Krichever78,Krichever80} and used in all
subsequent works on the subject is based on the well-known fact that
nonlinear integrable equations such as KP, BKP, CKP,
as well as the 2D Toda lattice can be
represented as compatibility conditions for systems of 
certain overdetermined 
linear problems which are partial differential or difference equations
in two variables (space and time). 
The suggested scheme of finding the dynamics of poles
consists in substituting the 
elliptic solution not in the nonlinear equation itself but in the 
linear problems for it, using a suitable pole ansatz for the 
wave function depending on a spectral parameter.

In this paper we systematically use another method 
suggested by one of the authors 
in \cite{K05} on the example of the KP/Calogero-Moser 
correspondence and further discussed in \cite{K10,K22}.
The key point of this method is existence of meromorphic solutions
to the linear partial differential or difference equations. 
We call them {\it monodromy free linear equations}. 
It turns out that 
the conditions
of existence of meromorphic solutions in the space variable 
are equivalent to equations
of motion for the poles which are equations of motion for
integrable many-body systems of Calogero-Moser and 
Ruijsenaars-Schneider type. 

We also prove that existence of at least one meromorphic solution 
implies existence of a whole family of meromorphic wave solutions 
depending on a spectral parameter. 

For completeness, we include in this paper the analysis of meromorphic
solutions 
to the linear problems for the KP and Toda lattice equations 
leading to the Calogero-Moser and Ruijsenaars-Schneider systems
respectively (this is contained in the earlier works \cite{K05,K10}). 
Conditions of existence of meromorphic
solutions to the linear problems for the BKP and CKP equations 
as well as for the Toda lattice of type C were not
discussed in the literature; in this paper 
we find them and show that they are equivalent to the
equations of motion for poles of these equations obtained
in \cite{RZ20,KZ20,KZ21a}.
 
The main new result of this paper is the equations of motion (\ref{b13})
for poles of elliptic solutions to the Toda lattice of type B 
recently introduced in \cite{KZ22}. They have the form
\beq\label{int1}
\ddot x_i +\sum_{k=1, \, \neq i}^N 
\dot x_i \dot x_k \Bigl (\zeta (x_{ik}+\eta )+
\zeta (x_{ik}-\eta )-2\zeta (x_{ik})\Bigr ) -U(x_{i1}, \ldots \, x_{iN})=0,
\eeq
where dot means the time derivative, $x_{ik}\equiv x_i-x_k$, 
\beq\label{int2}
U(x_{i1}, \ldots \, x_{iN})=\sigma (2\eta )\left [
\prod_{j\neq i}\frac{\sigma (x_{ij}+2\eta )\sigma (x_{ij}-\eta )}{\sigma
(x_{ij}+\eta )\sigma (x_{ij})}-
\prod_{j\neq i}\frac{\sigma (x_{ij}-2\eta )\sigma (x_{ij}+\eta )}{\sigma
(x_{ij}-\eta )\sigma (x_{ij})}\right ]
\eeq
and $\sigma (x)$, $\zeta (x)$ are the standard Weierstrass functions
(see Appendix A). 
These equations are obtained from the 
condition of existence of meromorphic solutions to the 
differential-difference auxiliary linear problem for the  
Toda lattice of type B. We also show that the same equations can be
obtained by restriction of the Ruijsenaars-Schneider dynamics 
with respect
to the time flow $\p_{t_1}-\p_{\bar t_1}$ of the system containing $2N$ particles to the half-dimensional 
subspace of the $4N$-dimensional phase space corresponding
to the configuration in which the particles stick together joining 
in pairs such that the distance between particles in each pair is equal to
$\eta$. This configuration is destroyed by the flow 
$\p_{t_1}+\p_{\bar t_1}$ but is preserved by the flow
$\p_{t_1}-\p_{\bar t_1}$ (and, hypothetically, by all higher flows 
$\p_{t_k}-\p_{\bar t_k}$) and the time evolution in $t=t_1-\bar t_1$  
of the pairs with coordinates $x_i$ is given by equations (\ref{int1}),
(\ref{int2}). This fact is not so surprising if we recall that 
the tau-function $\tau^{\rm Toda}(x)$ 
of the Toda lattice (whose zeros move as 
Ruijsenaars-Schneider particles) is connected with the tau-function 
$\tau (x)$ of the Toda lattice of type B (whose zeros move according
to equations (\ref{int1})) by the relation
\beq\label{int3}
\tau^{\rm Toda}(x)=\tau (x)\, \tau(x-\eta )
\eeq
(see \cite{KZ22}), so zeros of $\tau^{\rm Toda}(x)$ stick together in pairs.

To avoid a confusion, we should stress that what we mean by 
the Toda lattice of type B or C is very different from the systems
introduced in \cite{UT84} under similar names. Our equations are 
natural integrable discretizations of the BKP and CKP equations, that is
why we found it appropriate to call them ``Toda lattices of type B and C''.

The reader should be aware that the notation for coefficients of Laurent
expansions below are valid only throughout each section and the same notation may mean different things in different sections. 
We hope that this will not lead
to a misunderstanding since each section is devoted to its own 
linear equation and the contents do not
intersect. 

\section{Differential equations}

\subsection{The KP case}

We start with a worm-up exercise following \cite{K05}. 

Consider the linear equation
\beq\label{kp1}
(\p_t -\p_x^2 -2u)\psi =0,
\eeq
which is one of the auxiliary linear problems for the KP equation. 
It is easy to see that
if $u(x)$ has a pole $a$ in the complex $x$-plane, it must be a second order pole. Expanding 
the left hand side in a neighborhood of the pole, one can find a necessary condition of 
existence of a meromorphic solution in this neighborhood. 

\begin{proposition}
If equation (\ref{kp1}) with 
\beq\label{kp2}
u(x)=-\frac{1}{(x-a)^2} +u_0 +u_1 (x-a) + \ldots ,
\eeq
has a meromorphic in $x$ solution, then the condition
\beq\label{kp5}
\ddot a +4u_1=0
\eeq
holds, where dot means the time derivative.
\end{proposition}

\noindent
{\it Proof.}
Let the expansion of $\psi (x)$ around the point $a$ be of the form
\beq\label{kp3}
\psi (x) = \frac{\alpha}{x-a} +\beta +\gamma (x-a) +\delta (x-a)^2 +\ldots .
\eeq
Substituting expansions (\ref{kp2}), (\ref{kp3}) 
in the left hand side of (\ref{kp1}), we see that
the highest (third order) order poles cancel identically.  Equating the coefficients
in front of $(x-a)^{-2}$, $(x-a)^{-1}$ and $(x-a)^0$ to zero, we get the conditions
\beq\label{kp4}
\left \{\begin{array}{l}
\dot a \alpha +2\beta =0,
\\ \\
\dot \alpha +2\gamma -2u_0 \alpha =0,
\\ \\
\dot \beta -\dot a \gamma -2u_0 \beta -2u_1 \alpha =0,
\end{array}\right.
\eeq
Taking $t$-derivative of the first equation, plugging
$\dot \alpha$ and $\dot \beta$ from the second and the third ones 
and using the 
first one again, we obtain
the necessary condition (\ref{kp5}).
\square

One can see that this condition encodes equations of 
motion for the (elliptic in general)
Calogero-Moser system. Indeed, let $u(x)$ be the doubly-periodic meromorphic function
\beq\label{kp6}
u(x)=-\sum_i \wp (x-x_i),
\eeq
where $\wp (x)$ is the Weierstrass $\wp$-function, then the expansion (\ref{kp2}) near
the pole at $a=x_i$ holds true with
$$
u_0=-\sum_{j\neq i}\wp (x_i-x_j), \qquad 
u_1 = -\sum_{j\neq i}\wp ' (x_i-x_j),
$$
so the conditions (\ref{kp5}) for each $x_i$ read
\beq\label{kp7}
\ddot x_i =4\sum_{j\neq i}\wp '(x_i-x_j),
\eeq
which are the equations of motion for the elliptic Calogero-Moser system.

Next, we show that (\ref{kp5}) is simultaneously a sufficient condition for local existence 
of a meromorphic wave solution to equation (\ref{kp1}), 
i.e. a solution depending on a 
spectral parameter $k$ with the expansion of the form
\beq\label{kp8}
\psi (x) = e^{kx +k^2t}\Bigl (1+\sum_{s\geq 1} \xi_s k^{-s}\Bigr ), 
\quad k\to \infty .
\eeq

\begin{proposition}
Suppose that condition (\ref{kp5}) for the pole $a$ of $u(x)$ holds.
Then all wave solutions of equation (\ref{kp1})
of the form (\ref{kp8}) are meromorphic
in a neighborhood of the point $a$ with a simple pole
at $x=a$ and regular elsewhere in this neighborhood.
\end{proposition}

\noindent
{\it Proof.}
Substitution of the series (\ref{kp8}) 
into the equation (\ref{kp1}) gives the recurrence relation
\beq\label{kp9}
2\xi_{s+1}' =\dot \xi_s -2u\xi_s -\xi_s'', \qquad s\geq 0 , 
\qquad \xi_0\equiv 1.
\eeq
In particular, at $s=0$ we have
\beq\label{kp9a}
\xi_1'=-u.
\eeq
Let the Laurent expansion of $\xi_s$ near the pole at $x=a$ be of the form
\beq\label{kp10}
\xi_s = \frac{r_s}{x-a} +r_{s,0} +r_{s,1}(x-a) +\ldots ,
\eeq
and the expansion of $u(x)$ be as in (\ref{kp2}). 
The solution is meromorphic if the residue 
of the right hand side of (\ref{kp9})
vanishes:
\beq\label{kp11}
2\res_{x=a} \xi_{s+1}' =\dot r_s +2r_{s,1} -2u_0 r_s =0.
\eeq
At $s=0$ we have $\displaystyle{\res_{x=a}\xi_1'=0}$ from (\ref{kp9a}). We are going to prove 
(\ref{kp11}) by induction in $s$. 
Assume that (\ref{kp11}) holds for some $s$, then it is easy to 
see that the condition
(\ref{kp5}) implies that it holds for $s+1$. 
Indeed, substituting the expansion (\ref{kp10})
into the equation and equating the coefficients of $(x-a)^{-2}$, $(x-a)^{-1}$ and
$(x-a)^0$ to zero, we get the conditions
\beq\label{kp12}
\left \{
\begin{array}{l}
2r_{s+1}=-r_s\dot a -2r_{s,0},
\\ \\
\dot r_s +2r_{s,1} -2u_0 r_s =0,
\\ \\
2r_{s+1, 1}=\dot r_{s,0} -r_{s,1}\dot a -2u_0 r_{s,0}-2u_1 r_s.
\end{array}
\right.
\eeq
Substituting them into the right hand side of (\ref{kp11}) at $s\to s+1$, we have:
$$
2\res_{x=a} \xi_{s+2}' =\dot r_{s+1} +2r_{s+1,1} -2u_0 r_{s+1} 
$$
$$
\begin{array}{c}
=(-\frac{1}{2}\dot r_s \dot a -\frac{1}{2} r_s \ddot a -\dot r_{s,0}) +(\dot r_{s,0}
-r_{s,1}\dot a -2u_0 r_{s,0} -2u_1 r_s) +(u_0r_s\dot a +2u_0 r_{s,0}),
\end{array}
$$
where we have used the first and the third equations in (\ref{kp12}). After cancellations
we get:
$$
2\res_{x=a} \xi_{s+2}' =-\frac{1}{2} 
r_s (\ddot a +4u_1) -\frac{1}{2} \dot a
(\dot r_s +2r_{s,1} -2u_0 r_s)=0.
$$
(The second term vanishes by the induction assumption and the first one is zero
due to the condition (\ref{kp5}).)
\square

\subsection{The CKP case}

Consider the linear equation
\beq\label{ckp1}
(\p_t -\p_x^3 -6u\p_x -3u')\psi =0,
\eeq
which is one of the auxiliary linear problems for the CKP equation. 

\begin{proposition}
Suppose that $u(x)$ in (\ref{ckp1}) has a pole at $x=a$ and equation
(\ref{ckp1}) has a meromorphic solution, then the condition
\beq\label{ckp5}
\dot a + 6u_0=0
\eeq
holds, where $u_0$ is the coefficient in the Laurent expansion
\beq\label{ckp2}
u(x)=-\frac{1}{2(x-a)^2} +u_0 +u_1 (x-a) + \ldots .
\eeq
\end{proposition}

\noindent
{\it Proof.}
We have the expansion
near the pole at $x=a$:
\beq\label{ckp3}
\psi (x)= \frac{\alpha}{x-a} +\beta +\gamma (x-a) +\delta (x-a)^2 
+ \ldots .
\eeq
Substituting the expansions in 
the left hand side of (\ref{ckp1}), we see that
the highest (fourth order) order poles cancel identically.  The necessary 
condition of cancellation of the third order poles is $\beta =0$. Equating the coefficients
in front of $(x-a)^{-2}$, $(x-a)^{-1}$ and $(x-a)^0$ to zero, we get the conditions
\beq\label{ckp4}
\left \{\begin{array}{l}
\alpha \dot a  +6\alpha u_0 =0,
\\ \\
\dot \alpha +3\alpha u_1 +3\delta =0,
\\ \\
\gamma \dot a  +6\gamma u_0 =0.
\end{array}\right.
\eeq
The first and the third equations are equivalent and lead to the necessary condition (\ref{ckp5}).
\square

Let $u(x)$ be the elliptic function
\beq\label{ckp6}
u(x)=-\frac{1}{2}\sum_i \wp (x-x_i),
\eeq
then the expansion (\ref{ckp2}) near
the pole at $x=x_i$ holds true with
$\displaystyle{
u_0=-\frac{1}{2}\sum_{j\neq i}\wp (x_i-x_j), }
$
so the conditions (\ref{ckp5}) for each $x_i$ read
\beq\label{ckp7}
\dot x_i =3\sum_{j\neq i}\wp (x_i-x_j),
\eeq
which are the equations of motion for poles of elliptic solutions to the CKP equation
derived in \cite{KZ20}. As is shown in \cite{KZ20}, they are obtained by 
restriction of the third flow of the Calogero-Moser
system to the submanifold of turning points. 

Next, we show that (\ref{ckp5}) is simultaneously a sufficient condition for local existence 
of meromorphic wave solutions to equation (\ref{ckp1})
with the expansion of the form
\beq\label{ckp8}
\psi = e^{kx +k^3t}\Bigl (1+\sum_{s\geq 1} \xi_s k^{-s}\Bigr ), \quad k\to \infty .
\eeq

\begin{proposition}
Suppose that condition (\ref{ckp5}) for the pole $a$ of $u(x)$ holds.
Then all wave solutions of equation (\ref{ckp1})
of the form (\ref{ckp8})
are meromorphic in a neighborhood of the point $a$ 
with a simple pole at $x=a$ and regular elsewhere in this neighborhood.
\end{proposition}

\noindent
{\it Proof.}
Substitution of the series 
into equation (\ref{ckp1}) gives the recurrence relation
\beq\label{ckp9}
\dot \xi_s -3\xi_{s+2}' -3\xi_{s+1}'' -\xi_s'''
-6u\xi_{s+1} -6u\xi_s' -3u'\xi_s =0, \quad s\geq -1 , \quad \xi_{-1}\equiv 0, 
\; \xi_0\equiv 1.
\eeq
In particular, at $s=-1$ we have
\beq\label{ckp9a}
\xi_1'=-2u.
\eeq
It is convenient to represent equation (\ref{ckp9}) in the form
\beq\label{ckp9b}
f_s'=\dot \xi_s -6u\xi_{s+1} -3u\xi_s', \quad f_s=3\xi_{s+2}+3\xi_{s+1}' +\xi_s'' +3u\xi_s.
\eeq

Let the Laurent expansion of $\xi_s$ near the pole at $x=a$ be
\beq\label{ckp10}
\xi_s = \frac{r_s}{x-a} +r_{s,0} +r_{s,1}(x-a) +r_{s,2}(x-a)^2 +\ldots ,
\eeq
and the expansion of $u(x)$ be as in (\ref{ckp2}). 
The solution is meromorphic if the residue of the right hand side of (\ref{ckp9b})
vanishes:
\beq\label{ckp11}
\res_{x=a} f_{s}' =\dot r_s +3r_{s+1,1} +3r_{s,2}-6u_0 r_{s+1}+3u_1r_s =0.
\eeq
At $s=-1$ we have $\displaystyle{\res_{x=a}f_{-1}'=0}$ from (\ref{ckp9a}). We are going to prove 
(\ref{ckp11}) by induction in $s$. 
Assume that (\ref{ckp11}) holds for some $s$, then it is easy to see that the condition
(\ref{ckp5}) implies that it holds for $s+1$. 
Indeed, substituting the expansion (\ref{ckp10})
into the equation and equating the coefficients of $(x-a)^{-3}$, $(x-a)^{-2}$ 
$(x-a)^{-1}$ and
$(x-a)^0$ to zero, we get the conditions
\beq\label{ckp12}
\left \{
\begin{array}{l}
r_{s+1}+r_{s,0}=0,
\\ \\
3r_{s+2}+r_s\dot a +3r_{s+1,0}+6u_0r_s=0,
\\ \\
\dot r_s +3r_{s+1,1} +3r_{s,2}-6u_0 r_{s+1}+3u_1r_s =0,
\\ \\
\dot r_{s,0} -r_{s,1}\dot a -3r_{s+2, 1} -3r_{s+1,2}
-6u_0 r_{s+1,0}-6u_0r_{s,1}-3u_1 r_{s+1}=0.
\end{array}
\right.
\eeq
Substituting them into the right hand side 
of (\ref{ckp11}) at $s\to s+1$, we have:
$$
\res_{x=a} f_{s+1}' = \dot r_{s+1} +3r_{s+2,1} +
3r_{s+1,2}-6u_0 r_{s+2}+3u_1r_{s+1} 
$$
$$
=(2u_0r_s -r_{s,1})(\dot a+6u_0)-2u_0(3r_{s+2}+
r_s\dot a +3r_{s+1,0}+6u_0r_s)=0.
$$
(The second term vanishes because of the second equation in (\ref{ckp12}) and the first one
is zero 
due to the condition (\ref{ckp5}).)
\square

\subsection{The BKP case}

Consider the linear equation
\beq\label{bkp1}
(\p_t -\p_x^3 -6u\p_x)\psi =0,
\eeq
which is one of the auxiliary linear problems for the BKP equation. 

\begin{proposition}
Suppose that $u(x)$ in (\ref{bkp1}) has a pole at $x=a$ and equation
(\ref{bkp1}) has a meromorphic solution, then the condition
\beq\label{bkp5}
\ddot a +6\dot u_0 -12u_1(\dot a +6u_0) -36 u_3=0
\eeq
holds, where $u_0, u_1, u_3$ are coefficients in the Laurent expansion
\beq\label{bkp2}
u(x)=-\frac{1}{(x-a)^2} +u_0 +u_1 (x-a) + u_2 (x-a)^2 +u_3 (x-a)^3 +\ldots .
\eeq
\end{proposition}

\noindent
{\it Proof.}
We have the expansion
near the pole at $x=a$:
\beq\label{bkp3}
\psi (x)= 
\frac{\alpha}{x-a} +\beta +\gamma (x-a) +\delta (x-a)^2 +
\varepsilon (x-a)^3+ 
\mu (x-a)^4 +\ldots .
\eeq
Substituting the expansions in the left hand side of 
(\ref{bkp1}), we see that
possible fourth and third order poles cancel identically.  Equating the coefficients
in front of $(x-a)^{-2}$, $(x-a)^{-1}$, $(x-a)^0$ and $(x-a)$ 
to zero, we get the conditions
\beq\label{bkp4}
\left \{\begin{array}{l}
\alpha \dot a  +6\alpha u_0 +6\gamma =0,
\\ \\
\dot \alpha +6\alpha u_1 +12\delta =0,
\\ \\
\dot \beta -\gamma \dot a -6\gamma u_0 +6\alpha u_2 +12\varepsilon  =0,
\\ \\
\dot \gamma -2\delta \dot a -12\delta u_0 -6\gamma u_1 +6\alpha u_3 =0.
\end{array}\right.
\eeq
Note that the terms with the 
coefficient $\mu$ in (\ref{bkp3}) which might enter 
the last condition actually 
cancel. Taking $t$-derivative of the first equation, we have
$$
\ddot a +6\dot u_0 +6\frac{\dot \gamma}{\alpha}-6\frac{\gamma \dot \alpha}{\alpha^2}=0.
$$
Plugging here $\dot \alpha$ from the second equation and 
$\dot \gamma$ from the 
fourth one, we obtain the necessary condition (\ref{bkp5}).
\square

One can see that this condition encodes 
equations of motion for the 
many-body system obtained in \cite{RZ20} as a dynamical system for motion of poles of
elliptic solutions to the BKP equation. 
Indeed, let $u(x)$ be the doubly-periodic meromorphic function
\beq\label{bkp6}
u(x)=-\sum_i \wp (x-x_i),
\eeq
then the expansion (\ref{bkp2}) near
the pole at $x=x_i$ holds true with
$$
u_0=-\sum_{j\neq i}\wp (x_i-x_j), \quad
u_1=-\sum_{j\neq i}\wp ' (x_i-x_j), \quad
u_3=-\frac{1}{6}\sum_{j\neq i}\wp ''' (x_i-x_j)
$$
and
$\displaystyle{
\dot u_0 =-\sum_{j\neq i}(\dot x_i-\dot x_j)\wp '(x_i-x_j).}
$
Therefore, the conditions (\ref{bkp5}) for each $x_i$ give 
the equations of motion derived in \cite{RZ20}:
\beq\label{bkp7}
\ddot x_i+6 \sum_{j\neq i}(\dot x_i+\dot x_j)\wp '(x_i-x_j)-
72 \sum_{j\neq k\neq i}
\wp (x_i-x_j)\wp '(x_i-x_k)=0
\eeq
(the identity $\wp '''(x)=12\wp (x)\wp '(x)$ is used). 
As is shown in \cite{Z21},
the same equations can be obtained by 
restriction of the third flow of the elliptic Calogero-Moser
system to the subspace of the phase space in which $2N$ particles
stick together in pairs in such a way that the two particles in each
pair are in one and the same point. This configuration 
is immediately destroyed by the second
flow of the Calogero-Moser system but is preserved by the third one,
and coordinates of the pairs are subject to equations (\ref{bkp7}).

Next, we will show 
that (\ref{bkp5}) is simultaneously a sufficient condition for local 
existence 
of meromorphic wave solutions to equation (\ref{bkp1})
of the form
\beq\label{bkp8}
\psi (x) = e^{kx +k^3t}\Bigl (1+\sum_{s\geq 1} \xi_s k^{-s}\Bigr ), 
\quad k\to \infty .
\eeq

\begin{proposition}
Suppose that condition (\ref{bkp5}) for the pole of $u(x)$ holds.
Then all wave solutions of equation (\ref{bkp1})
of the form (\ref{bkp8}) are meromorphic
in a neighborhood of the point $a$ with
a simple pole at $x=a$ and regular elsewhere in this neighborhood.
\end{proposition}

\noindent
{\it Proof.}
Substitution of the series into the equation 
(\ref{bkp1}) gives the recurrence relation
\beq\label{bkp9}
\dot \xi_s -3\xi_{s+2}' -3\xi_{s+1}'' -\xi_s'''
-6u\xi_{s+1} -6u\xi_s' =0, \quad s\geq -1 , \quad \xi_{-1}\equiv 0, 
\; \xi_0\equiv 1.
\eeq
In particular, at $s=-1$ we have
\beq\label{bkp9a}
\xi_1'=-2u.
\eeq
It is convenient to represent equation (\ref{bkp9}) in the form
\beq\label{bkp9b}
g_s'=\dot \xi_s -6u\xi_{s+1} -6u\xi_s', \quad g_s=3\xi_{s+2}+3\xi_{s+1}' +\xi_s''.
\eeq

Let the Laurent expansion of $\xi_s$ near the pole at $x=a$ be
\beq\label{bkp10}
\xi_s = \frac{r_s}{x-a} +r_{s,0} +r_{s,1}(x-a) +
r_{s,2}(x-a)^2 +r_{s,3}(x-a)^3 +\ldots ,
\eeq
and the expansion of $u(x)$ be as in (\ref{bkp2}). 
The solution is meromorphic if the residue of the right hand side of (\ref{bkp9b})
vanishes:
\beq\label{bkp11}
\res_{x=a} g_{s}' =\dot r_s +6r_{s+1,1} +12r_{s,2}-6u_0 r_{s+1}+6u_1r_s =0.
\eeq
At $s=-1$ we have $\displaystyle{\res_{x=a}g_{-1}'=0}$ from (\ref{bkp9a}). 
As before, we will prove 
(\ref{bkp11}) by induction in $s$. 
Assume that (\ref{bkp11}) holds for some $s$, 
then it is easy to see that the condition
(\ref{bkp5}) implies that it holds for $s+1$. 
Indeed, substituting the expansion (\ref{bkp10})
into the equation and equating the coefficients of $(x-a)^{-2}$, $(x-a)^{-1}$ 
$(x-a)^{0}$ and
$(x-a)$ to zero, we get the conditions
\beq\label{bkp12}
\left \{
\begin{array}{l}
3r_{s+2}+r_s\dot a +6r_{s+1,0}+6r_{s,1}+6u_0r_s=0,
\\ \\
\dot r_s +6r_{s+1,1} +12r_{s,2}-6u_0 r_{s+1}+6u_1r_s =0,
\\ \\
\dot r_{s,0} -r_{s,1}\dot a -3r_{s+2, 1} +12r_{s,3}
-6u_0 r_{s+1,0}-6u_0r_{s,1}-6u_1 r_{s+1}+6u_2r_s=0,
\\ \\
\begin{array}{l}
\dot r_{s,1} -2r_{s,2}\dot a -6r_{s+2, 2} -12r_{s+1,3}
-6u_0 r_{s+1,1}-12u_0r_{s,2}
\\ \\
\phantom{aaaaaaaaaaaaaaaaaaaaaaa}-6u_1 r_{s+1,0}-6u_1r_{s,1}-6u_2r_{s+1}+6u_3r_s=0.\end{array}
\end{array}
\right.
\eeq
Substituting them into the right hand side of (\ref{bkp11}) at $s\to s+1$, we have:
$$
3\res_{x=a} g_{s+1}' = 3\dot r_{s+1} +18r_{s+2,1} +36r_{s+1,2}-18u_0 r_{s+2}+18u_1r_{s+1} 
$$
$$
=-r_{s-1}(\ddot a +6\dot u_0 -12u_1(\dot a+6u_0)-36u_3)-
\dot a(\dot r_{s-1}+12r_{s-1,2}+6r_{s,1}-6u_0r_s +6u_1r_{s-1})
$$
$$
-6u_0(r_s\dot a +3r_{s+2}+6r_{s+1,0}+6r_{s,1}+6u_0r_s )
-6u_1(r_{s-1}\dot a +3r_{s+1}+6r_{s,0}+6r_{s-1,1}+6u_0r_{s-1} )=0.
$$
(The last two terms vanish because of the first equation in (\ref{bkp12}),
the second term vanishes due to the induction assumption and the first one
is zero 
due to the condition (\ref{bkp5}).)
\square

\section{Differential-difference equations}

\subsection{The Toda lattice case}

The 2D Toda lattice equation is the compatibility condition for 
the linear differential-difference equations 
\beq\label{t1}
\p_{t_1} \psi (x)=\psi (x+\eta ) +b(x)\psi (x),
\eeq
\beq\label{t2}
\p_{\bar t_1}\psi (x) =v(x)\psi (x-\eta ),
\eeq
where $\eta$ is a parameter (the lattice spacing). 
Suppose that $b(x)$ and $v(x)$ are meromorphic functions;
let us investigate when these equations have meromorphic solutions
in $x$. 

\subsubsection{The linear problem with respect to $t_1$}

First we consider the linear problem (\ref{t1}):
\beq\label{t3}
\p_{t} \psi (x)=\psi (x+\eta ) +b(x)\psi (x).
\eeq
Let $b(x)$ have a first order pole at $x=a$, then the equation 
requires that it has also
a pole at $x=a-\eta$:
\beq\label{t4}
b(x)=\left \{ 
\begin{array}{l}
\displaystyle{ \phantom{-}
\frac{\nu}{x-a}\, +\, \mu_0 \, + \, O(x-a), \quad x\to a}
\\ \\
\displaystyle{-\frac{\nu}{x-a+\eta}+\mu_1 +O(x-a+\eta ), \quad x\to a-\eta .}
\end{array}
\right.
\eeq

\begin{proposition}
Suppose that $b(x)$ in (\ref{t3}) has poles at $x=a$ and 
$x=a-\eta$ with expansions near the poles 
of the form (\ref{t4}). If equation
(\ref{t3}) has a meromorphic solution with a pole at the point $a$
regular at $a\pm \eta$, then the condition
\beq\label{t10}
\ddot a -\dot a (\mu_0 +\mu_1)=0
\eeq
holds.
\end{proposition}

\noindent
{\it Proof.}
Let the expansion of $\psi (x)$ around the point $a$ be of the form
\beq\label{t5}
\psi (x)=\frac{\alpha}{x-a} +\beta +O(x-a),
\eeq
then
$$
\p_t \psi (x)=\frac{\alpha \dot a}{(x-a)^2}+\frac{\dot \alpha}{x-a}+
O(1).
$$
Substituting the expansions around the point $a$ in the equation, 
we write:
$$
\frac{\alpha \dot a}{(x-a)^2}+\frac{\dot \alpha}{x-a}+
O(1)=\left (\frac{\nu}{x-a}\, +\, \mu_0 +\ldots \right )
\left (\frac{\alpha}{x-a} +\beta +\ldots \right ).
$$
Equating the coefficients in front of the poles, we obtain the 
conditions
\beq\label{t6}
\left \{ \begin{array}{l}
\nu =\dot a,
\\ \\
\dot \alpha =\nu \beta +\mu_0 \alpha .
\end{array}
\right.
\eeq
Around the point $a-\eta$ we have:
$$
\begin{array}{c}
\p_t \psi (a-\eta ) +O(x-a+\eta ) 
\\ \\
\displaystyle{=\frac{\alpha}{x-a+\eta} +\beta 
-\Bigl (\frac{\nu}{x-a+\eta} +\mu_1 +\ldots \Bigr )
\Bigl (\psi (a-\eta ) +(x-a+\eta )\psi '(a-\eta )+\ldots \Bigr ).}
\end{array}
$$
Equating the coefficients in front of the terms or order $(x-a+\eta )^{-1}$
and $(x-a+\eta )^{0}$, we obtain the 
conditions
\beq\label{t7}
\left \{ \begin{array}{l}
\alpha =\nu \psi (a-\eta ),
\\ \\
\p_t \psi (a-\eta )=\beta -\mu_1 \psi (a-\eta )-\nu \psi '(a-\eta ).
\end{array}
\right.
\eeq
Taking the time derivative of the first equation in (\ref{t7})
and using (\ref{t6}), we 
get
\beq\label{t8}
\dot \alpha = \ddot a \psi (a-\eta )+\dot a \dot \psi (a-\eta ),
\eeq
where
\beq\label{t9} 
\dot \psi (a-\eta )=\p_t \psi (a-\eta )+\dot a \psi '(a-\eta )
\eeq
is the full time derivative of $\psi (a-\eta )$. Now, combining
equations (\ref{t6})--(\ref{t9}), we arrive at the condition (\ref{t10}).
\square

Suppose now that $b(x)$ is an elliptic function 
of $x$ having $2N$ first order poles in the fundamental domain at some
points $x_j$ and at the points $x_j-\eta$, then it must have the form
\beq\label{t11}
b(x)=\sum_j \dot x_j \Bigl (\zeta (x-x_j)-\zeta (x-x_j+\eta )\Bigr ).
\eeq
Let $a=x_i$ for some $i$, then the coefficients $\mu_0$, $\mu_1$ in
(\ref{t4}) are
\beq\label{t12}
\begin{array}{l}
\displaystyle{
\mu_0 =\sum_{k\neq i} \dot x_k \zeta (x_i-x_k)-\sum_k \dot x_k
\zeta (x_i-x_k +\eta ),}
\\ \\
\displaystyle{
\mu_1 =\sum_{k\neq i} \dot x_k \zeta (x_i-x_k)-\sum_k \dot x_k
\zeta (x_i-x_k -\eta )}
\end{array}
\eeq
and (\ref{t10}) is equivalent to the equation of motion
\beq\label{t13}
\ddot x_i +\sum_{k\neq i}\dot x_i \dot x_k \Bigl (
\zeta (x_i-x_k+\eta )+\zeta (x_i-x_k-\eta )-2\zeta (x_i-x_k)\Bigr )=0
\eeq
of the Ruijsenaars-Schneider model.

Let us show that
(\ref{t10}) is a sufficient condition for existence of a meromorphic
wave solution to equation (\ref{t3}) depending on a spectral parameter $k$
with a pole at $x=a$ and regular 
at the points $x=a\pm \eta$. This solution can be found in the form
\beq\label{t14}
\psi (x)= k^{x/\eta}e^{kt} \Bigl (1+ \sum_{s\geq 1} \xi_s(x)k^{-s}\Bigr ),
\quad k\to \infty .
\eeq

\begin{proposition}
Suppose that $b(x)$ in (\ref{t3}) has poles at $x=a$ and 
$x=a-\eta$ with expansions near the poles 
of the form (\ref{t4}) and
condition (\ref{t10}) for the pole of $b(x)$ holds.
Then all wave solutions of equation (\ref{t3})
of the form (\ref{t14}) are meromorphic in a neighborhood
of the point $a$ with
a simple pole at $x=a$ and regular at
$x=a\pm \eta$.
\end{proposition}

\noindent
{\it Proof.}
Substituting the expansions into the equation, 
we obtain the recurrence relation
for the coefficients $\xi_s$:
\beq\label{t15}
\xi_{s+1}(x)-\xi_{s+1}(x+\eta )=b(x)\xi_s(x)-\dot \xi_s (x), \quad s\geq 0
\eeq
(it is convenient to put $\xi_0=1$). 
In particular, 
\beq\label{t16}
\xi_{1}(x)-\xi_{1}(x+\eta )=b(x).
\eeq
Let the Laurent expansion of $\xi_s$ near the point $a$ be
\beq\label{t17}
\xi_s(x)=\frac{r_s}{x-a} +r_{s,0}+r_{s,1}(x-a) +\ldots \, , 
\quad x\to a.
\eeq
Substituting this into (\ref{t15}) with $x\to a$ 
and equating the coefficients at the
poles, we get the recurrence relation
\beq\label{t18}
r_{s+1}=\dot a r_{s,0} +\mu_0 r_s -\dot r_s.
\eeq
Expanding (\ref{t15}) near the point $x\to a-\eta$ and equating the
coefficients in front of $(x-a+\eta )^{-1}$ and $(x-a+\eta )^{0}$, we get
\beq\label{t19}
\left \{ \begin{array}{l}
r_{s+1}=\dot a \xi_s(a-\eta ),
\\ \\
r_{s+1, 0}-\xi_{s+1}(a-\eta )=\mu_1 \xi_s(a-\eta )+\dot a \xi_s'(a-\eta )
+\p_t \xi_s (a-\eta ).
\end{array}
\right.
\eeq
The meromorphic wave solution with the required properties exists
if the sum of residues of the right hand side of (\ref{t15}) at the 
points $x=a$ and $x=a-\eta$ is equal to zero for all $s\geq 0$. 
This sum of residues
is given by
$$
R_s= \dot a r_{s,0} +\mu_0 r_s -\dot r_s -\dot a \xi_s(a-\eta ).
$$
We are going to prove that $R_s=0$ for all $s\geq 0$ by induction. 
This is obviously true for $s=0$ due to equation (\ref{t16}). 
Assume that $R_s=0$ for some $s$, this is our induction assumption.
Using (\ref{t19}), we find:
$$
R_{s+1}=\dot a \Bigl (r_{s+1,0}-\xi_{s+1}(a-\eta )\Bigr ) 
+\mu_0 r_{s+1} -\dot r_{s+1} 
$$
$$
=\Bigl (\dot a(\mu_0+\mu_1)-\ddot a\Bigr )\xi_s(a-\eta )-\p_t R_s=0
$$
because the first term vanishes due to the condition (\ref{t10}) 
and the second one vanishes due to the induction assumption. 
\square

\subsubsection{The linear problem with respect to $\bar t_1$}

Consider now the linear problem (\ref{t2}):
\beq\label{bar1}
\p_{t}\psi (x) =v(x)\psi (x-\eta ).
\eeq
Suppose that 
the function $v(x)$ has a second order pole at $x=a$ with the
expansion
\beq\label{bar2}
v(x)=\frac{\nu}{(x-a)^2} +\frac{\mu}{x-a}+O(1), \quad x\to a .
\eeq
We also assume that $v(x)$ has a zero at $x=a-\eta$:
$
v(a-\eta )=0.
$

\begin{proposition}
Suppose that $v(x)$ in (\ref{bar1}) has a second order pole at $x=a$ 
with the expansion (\ref{bar2}) and $v(a-\eta )=0.$
If equation
(\ref{bar1}) has a meromorphic solution with a simple 
pole at the point $a$
regular at $a\pm \eta$, then the condition
\beq\label{bar6}
\nu \ddot a+\mu \dot a^2 -\dot \nu \dot a=0
\eeq
holds.
\end{proposition}

\noindent
{\it Proof.}
For $\psi (x)$ near the point $a$ we have
\beq\label{bar3}
\psi (x)=\frac{\alpha}{x-a}+O(1), \quad x\to a.
\eeq
Substituting the expansions around the point $a$ in 
the equation (\ref{bar1}), 
we write:
$$
\frac{\alpha \dot a}{(x-a)^2}+\frac{\dot \alpha}{x-a}
=\Bigl (\frac{\nu}{(x-a)^2} +\frac{\mu}{x-a}+O(1)\Bigr )
\Bigl (\psi (a-\eta )+(x-a)\psi '(a-\eta )+\ldots \Bigr ).
$$
Equating the coefficients in front of the poles, we obtain the 
conditions
\beq\label{bar4}
\left \{ \begin{array}{l}
\alpha \dot a =\nu \psi (a-\eta ),
\\ \\
\dot \alpha =\nu \psi '(a-\eta ) +\mu \psi (a-\eta ) .
\end{array}
\right.
\eeq
At $x=a-\eta$ equation (\ref{bar1}) gives
$
\p_t \psi (a-\eta )=0
$
due to the fact that 
$v(a-\eta )=0$, so the full time derivative of $\psi (a-\eta )$ is
\beq\label{bar5}
\dot \psi (a-\eta )=\dot a \psi '(a-\eta ).
\eeq
Combining the time derivative of the first equation in (\ref{bar4}) 
with the second one and
using (\ref{bar5}), we obtain the condition (\ref{bar6}).
\square

Suppose that $v(x)$ is an elliptic function of the form
\beq\label{bar7}
v(x)=\prod_{j=1}^N \frac{\sigma (x-x_j+\eta )
\sigma (x-x_j-\eta )}{\sigma ^2
(x-x_j)},
\eeq
then, setting $a=x_i$, we have:
\beq\label{bar8}
\nu = -\sigma^2(\eta )\prod_{j\neq i}
\frac{\sigma (x_i-x_j+\eta )\sigma (x_i-x_j-\eta )}{\sigma ^2
(x_i-x_j)},
\eeq
\beq\label{bar9}
\mu =\nu \sum_{k\neq i}\Bigl (
\zeta (x_i-x_k+\eta )+\zeta (x_i-x_k-\eta )-2\zeta (x_i-x_k)\Bigr ),
\eeq
and the condition (\ref{bar6}) is equivalent to the equation of motion
which is the same as (\ref{t13}). 

Similarly to the previous case, equation
(\ref{bar6}) is a sufficient condition for local 
existence of a meromorphic
wave solution to equation (\ref{bar1}) depending on a spectral parameter $k$
with a pole at $x=a$ and regular 
at the points $x=a\pm \eta$. 
However, the arguments need some modifications. 

\begin{proposition}
Suppose that $v(x)$ in (\ref{bar1}) has a second order pole at $x=a$ 
with expansion near the pole
of the form (\ref{bar2}) and $v(a-\eta )=0$. 
Let $v(x)$ be of the form
\beq\label{bar10}
v(x)=e^{\varphi (x)-\varphi (x-\eta )},
\eeq
where $\varphi (x)$ has logarithmic singularities at $x=a$ and
$x=a-\eta$, so that
\beq\label{bar11}
\dot \varphi (x)=\left \{ 
\begin{array}{l}
\displaystyle{\frac{\dot a}{x-a} +\varphi_0 +O(x-a)}, \quad x\to a,
\\ \\
\displaystyle{-\frac{\dot a}{x-a+\eta} +\varphi_1 +O(x-a+\eta )}, 
\quad x\to a-\eta .
\end{array}
\right.
\eeq
Suppose that
condition (\ref{bar6}) for the pole of $v(x)$ holds.
Then all wave solutions of equation (\ref{bar1})
of the form 
\beq\label{bar13}
\psi (x) = k^{-x/\eta}e^{kt +\varphi (x)} 
\Bigl (1+ \sum_{s\geq 1} \xi_s(x)k^{-s}\Bigr ),
\quad k\to \infty 
\eeq
are meromorphic in a neighborhood of the point $a$
with a simple pole at $x=a$ and regular at
$x=a\pm \eta$.
\end{proposition}

\noindent
{\it Proof.}
Expanding the equality $\p_t \log v(x)=\dot 
\varphi (x)-\dot \varphi (x-\eta )$ near the point $x=a$ with the help
of (\ref{bar2}) and comparing the coefficients, we get:
\beq\label{bar12}
\varphi_0-\varphi_1 =\frac{\dot \nu}{\nu}-\frac{\mu}{\nu}\, \dot a.
\eeq
Substituting the expansions into the equation, 
we obtain the recurrence relation
for the coefficients $\xi_s$:
\beq\label{bar14}
\xi_{s+1}(x-\eta )-\xi_{s+1}(x)=
\dot \varphi (x)\xi_s(x)+\dot \xi_s (x), \quad s\geq 0.
\eeq
In particular, 
\beq\label{bar15}
\xi_{1}(x-\eta )-\xi_{1}(x)=\dot \varphi (x).
\eeq
The factor $e^{\varphi (x)}$ in (\ref{bar13}) has a pole at $x=a$ and
a zero at $x=a-\eta$. This means that the coefficients $\xi_s(x)$ 
are regular at $x=a$ and may have a pole at $x=a-\eta$.
Let the Laurent expansion of $\xi_s$ near the point $a-\eta$ be of the form
\beq\label{bar16}
\xi_s(x)=\frac{r_s}{x-a+\eta} +r_{s,0}+r_{s,1}(x-a+\eta ) +\ldots \, , 
\quad x\to a-\eta .
\eeq
The meromorphic wave solution with the required properties exists
if the sum of residues of the right hand side of (\ref{bar14}) at the 
points $x=a$ and $x=a-\eta$ is equal to zero for all $s\geq 0$. 
This can be proved by induction in the same way as for the linear problem
(\ref{t3}). 
\square

\subsection{The case of the Toda lattice of type C}

The Toda lattice of type C was introduced in the paper \cite{KZ21a}
by the authors. The auxiliary linear problem has the form
\beq\label{c1}
\p_t \psi (x)=\psi (x+\eta )+\frac{1}{2}\, \dot \varphi (x)\psi (x)+
v(x)\psi (x-\eta ),
\eeq
where
$
v(x)=e^{\varphi (x)-\varphi (x-\eta )}.
$
We assume that $\dot \varphi (x)$ is expanded near the points
$x=a$ and $x=a-\eta$ as in (\ref{bar11}) 
and
\beq\label{c2}
v(x)=\frac{\nu}{(x-a)^2}+\frac{\mu}{x-a}+O(1), \quad x\to a.
\eeq
We also note that the relation (\ref{bar12}) holds. 

\begin{proposition}
Suppose that $v(x)=e^{\varphi (x)-\varphi (x-\eta )}$ 
in (\ref{c1}) has a second order pole at $x=a$ 
with the expansion (\ref{c2}), $v(a-\eta )=0$ and
$\dot \varphi (x)$ is expanded near the points
$x=a$ and $x=a-\eta$ as in (\ref{bar11}).
If equation
(\ref{c1}) has a meromorphic solution with a simple 
pole at the point $a$
regular at $a\pm \eta$, then the condition
\beq\label{c5}
\dot a^2=-4\nu
\eeq
holds.
\end{proposition}

\noindent
{\it Proof.}
Let the function
$\psi (x)$ have a pole at $x=a$ with the expansion near this point
of the form (\ref{t5}). Substituting the expansions
near $x=a$ into the equation and equating the coefficients in front
of the highest poles, we get the condition
\beq\label{c3}
\alpha \dot a + 2\nu \psi (a-\eta )=0.
\eeq
The same procedure at the pole at $x=a-\eta$ leads to the condition
\beq\label{c4}
2\alpha =\dot a \psi (a-\eta ).
\eeq
Combining (\ref{c3}) and (\ref{c4}), we obtain the condition (\ref{c5}).
\square

Suppose now that $v(x)$ is the elliptic function (\ref{bar7}), then,
expanding it near the point $a=x_i$, we see that
$\nu$ is given by (\ref{bar8}). Equations (\ref{c5}) for any $i$ 
are then 
equivalent to the
equations of motion for poles of elliptic solutions to the Toda 
lattice of type C:
\beq\label{c6}
\dot x_i = 2\sigma (\eta )\prod_{j\neq i}
\frac{(\sigma (x_i-x_j+\eta )\sigma (x_i-x_j-\eta ))^{1/2}}{\sigma
(x_i-x_j)}.
\eeq
These equations were obtained in \cite{KZ21a}. The properly taken limit
$\eta \to 0$ leads to equations (\ref{ckp7}). 

Let us show that
(\ref{c5}) is a sufficient condition for existence of a meromorphic
wave solution to equation (\ref{c1}) depending on a spectral parameter $k$
with a pole at $x=a$ and regular 
at the points $x=a\pm \eta$.

\begin{proposition}
Suppose that $v(x)=e^{\varphi (x)-\varphi (x-\eta )}$ 
in (\ref{c1}) has a second order pole at $x=a$ 
with expansion near the pole
of the form (\ref{c2}) and $v(a-\eta )=0$. 
Assume also that the function $\dot \varphi (x)$ has expansion
of the form (\ref{bar11}).
If
condition (\ref{bar6}) for the pole of $v(x)$ holds, 
then all wave solutions of equation (\ref{c1})
of the form 
\beq\label{c7}
\psi (x)= k^{x/\eta}e^{kt} 
\Bigl (1+ \sum_{s\geq 1} \xi_s(x)k^{-s}\Bigr ),
\quad k\to \infty 
\eeq
are meromorphic in a neighborhood of the point $a$ with
a simple pole at $x=a$ and regular at
$x=a\pm \eta$.
\end{proposition}

\noindent
{\it Proof.}
Substituting the expansions into the equation, 
we obtain the recurrence relation
for the coefficients $\xi_s$ in (\ref{c7}):
\beq\label{c8}
\xi_{s+1}(x)-\xi_{s+1}(x+\eta )=
\frac{1}{2}\, \dot \varphi (x)\xi_s(x)-\dot \xi_s (x)
-v(x)\xi_{s-1}(x-\eta ), \quad s\geq 0,
\eeq
where we set $\xi_{-1}=0$, $\xi_0=1$. 
In particular, 
\beq\label{c9}
\xi_{1}(x)-\xi_{1}(x+\eta )=\frac{1}{2}\, \dot \varphi (x).
\eeq
Assuming the expansion of $\xi_s(x)$ near the point $a$ of the form
(\ref{t17}), we get from (\ref{c8}) expanded near the this point:
\beq\label{c10}
\left \{ \begin{array}{l}
\frac{1}{2}\, r_s \dot a +\nu \xi_{s-1}(a-\eta )=0,
\\ \\
r_{s+1}=\frac{1}{2}\, \dot a r_{s,0} +\frac{1}{2}\, \varphi_0 r_s -
\dot r_s -\nu \xi_{s-1}'(a-\eta )-\mu \xi_{s-1}(a-\eta ).
\end{array}
\right.
\eeq
The expansion near the point $x=a-\eta$ gives:
\beq\label{c11}
\left \{ \begin{array}{l}
r_{s+1}=\frac{1}{2}\, \dot a \xi_s(a-\eta ),
\\ \\
\xi_{s+1}(a-\eta )-r_{s+1,0}=\frac{1}{2}\, \varphi_1 \xi_s(a-\eta )-
\frac{1}{2}\, \dot a \xi_{s}'(a-\eta )-\p_t\xi_{s}(a-\eta ).
\end{array}
\right.
\eeq
Let 
$$
R_s=\frac{1}{2}\, \dot a (r_{s,0}-\xi_s(a-\eta ))+\frac{1}{2}\, \varphi_0
r_s -\dot r_s -\nu \xi_{s-1}'(a-\eta )-\mu \xi_{s-1}(a-\eta )
$$
be sum of the residues at the points $x=a$ and $x=a-\eta$ which 
must be zero. At $s=0$ this is true due to (\ref{c9}). Our induction
assumption is that $R_s=0$ for some $s$; let us show that this implies
that $R_{s+1}=0$. We have:
$$
R_{s+1}=\frac{1}{2}\, \dot a (r_{s+1,0}-\xi_{s+1}
(a-\eta ))+\frac{1}{2}\, \varphi_0
r_{s+1} -\dot r_{s+1} -\nu \xi_{s}'(a-\eta )-\mu \xi_{s}(a-\eta ).
$$
Substituting here the recurrence relations
(\ref{c10}), (\ref{c11}), we obtain, after some 
calculations and cancellations: 
$$
R_{s+1}=
\frac{1}{4}\Bigl ( (\varphi_0\! -\! \varphi_1)\dot a -4\mu -2\ddot a\Bigr )
\xi_s(a-\eta ) 
-\frac{1}{4}\, (\dot a^2+4\nu )\xi_s'(a-\eta )-\p_t R_s.
$$
The last two terms are equal to zero by virtue of the induction
assumption and the condition (\ref{c5}). As for the first term, we have:
$$
(\varphi_0\! -\! \varphi_1)\dot a -4\mu -2\ddot a=
\Bigl (\frac{\dot \nu}{\nu}-\frac{\mu}{\nu}\, \dot a \Bigr )\dot a
-4\mu -
2\ddot a
$$
$$
=(\dot a^2 +4\nu )\frac{\dot \nu -\mu \dot a}{\nu \dot a}-
\frac{1}{\dot a}\, \p_t(\dot a^2 +4\nu )=0
$$
by virtue of the condition (\ref{c5}). Therefore, $R_{s+1}=0$ and
we have proved the existence of a meromorphic wave solution.
\square

\subsection{The case of the Toda lattice of type B}

\subsubsection{Existence of a meromorphic solution}

The Toda lattice of type B was recently introduced by the authors 
in \cite{KZ22}.
The linear equation for the first time flow
has the form
\beq\label{b1}
\p_t \psi (x)=v(x)(\psi (x+\eta )-\psi (x-\eta )).
\eeq
Assume that the function $v(x)$ has a second order pole at $x=a$ with the
expansion
\beq\label{b2}
v(x)=\frac{\nu}{(x-a)^2} +\frac{\mu}{x-a}+O(1), \quad x\to a .
\eeq
We also assume that $v(x)$ has zeros at $x=a-\eta$ and $x=a+\eta$:
\beq\label{b3}
v(x)=\left \{
\begin{array}{l}
(x-a-\eta )V^+(a)+O((x\! -\! a\! -\! \eta )^2), \quad x\to a+\eta ,
\\ \\
(x-a+\eta )V^-(a)+O((x\! -\! a\! +\! \eta )^2), \quad x\to a-\eta .
\end{array}\right.
\eeq

\begin{proposition}
Suppose that $v(x)$ in (\ref{b1}) has a second order pole at $x=a$ 
and zeros at $x=a\pm \eta$
with the expansions (\ref{b2}), (\ref{b3}).  
If equation
(\ref{b1}) has a meromorphic solution with a simple 
pole at the point $a$
regular at $a\pm \eta$, then the condition
\beq\label{b8}
\nu \ddot a +\mu \dot a^2 -\dot \nu \dot a +\nu ^2 (V^+(a)+V^-(a))=0
\eeq
holds.
\end{proposition}

\noindent
{\it Proof.}
The principal part of the Laurent 
expansion of $\psi (x)$ near the point $x=a$ is
\beq\label{b4}
\psi (x)=\frac{\alpha}{x-a}+O(1), \quad x\to a.
\eeq
As $x\to a$, we have from equation (\ref{b1}):
$$
\frac{\alpha \dot a}{(x-a)^2}+\frac{\dot \alpha}{x-a}
$$
$$
=\Bigl (\frac{\nu}{(x-a)^2} \! +\! \frac{\mu}{x-a}\! +\! O(1)\Bigr )
\Bigl (\psi (a\! +\! \eta )\! -\! \psi (a\! -\! \eta )\! + \!(x-a)
(\psi '(a\! +\! \eta )\! -\! \psi '(a\! -\! \eta ))\! 
+\ldots \Bigr ).
$$
Equating the coefficients in front of the poles, we obtain the 
conditions
\beq\label{b5}
\left \{ \begin{array}{l}
\alpha \dot a =\nu (\psi (a+\eta )-\psi (a-\eta )),
\\ \\
\dot \alpha =\mu (\psi (a+\eta ) - \psi (a-\eta ))+\nu
(\psi '(a+\eta ) -\psi '(a-\eta ) ).
\end{array}
\right.
\eeq
At $x=a\pm \eta$ equation (\ref{b1}) gives:
\beq\label{b6}
\p_t \psi (a\pm \eta )=\mp \alpha \, V^{\pm}(a).
\eeq
Therefore,
\beq\label{b7}
\dot \psi (a\pm \eta )=\mp \alpha \, V^{\pm}(a)+\dot a \psi '(a\pm \eta ).
\eeq
Taking the time 
derivative of the first equation in (\ref{b5}) and combining it with the
other equations, we obtain the condition (\ref{b8}).
\square

\subsubsection{Dynamics of poles of elliptic solutions}

Suppose that $v(x)$ is an elliptic function of the form
\beq\label{b9}
v(x)=\prod_{j=1}^N \frac{\sigma (x-x_j+\eta )\sigma (x-x_j-\eta )}{\sigma ^2
(x-x_j)},
\eeq
then, setting $a=x_i$, we have:
\beq\label{b10}
\nu = -\sigma^2(\eta )\prod_{j\neq i}
\frac{\sigma (x_i-x_j+\eta )\sigma (x_i-x_j-\eta )}{\sigma ^2
(x_i-x_j)},
\eeq
\beq\label{b11}
\mu =\nu \sum_{k\neq i}\Bigl (
\zeta (x_i-x_k+\eta )+\zeta (x_i-x_k-\eta )-2\zeta (x_i-x_k)\Bigr ),
\eeq
\beq\label{b12}
V^{\pm}(x_i)=\pm \frac{\sigma (2\eta )}{\sigma^2(\eta )}
\prod_{j\neq i}\frac{\sigma (x_i-x_j\pm 2\eta )\sigma (x_i-x_j)}{\sigma^2
(x_i-x_j\pm \eta )},
\eeq
and the condition (\ref{b8}) is equivalent to the equation of motion
\beq\label{b13}
\begin{array}{c}
\displaystyle{
\ddot x_i +\sum_{k\neq i}\dot x_i \dot x_k \Bigl (
\zeta (x_i\! -\! x_k\! +\! \eta )+\zeta (x_i\! -\! x_k\! -\! \eta )
-2\zeta (x_i-x_k)\Bigr )}
\\ \\
\displaystyle{
-\sigma (2\eta )\left [ \prod_{j\neq i}
\frac{\sigma (x_i\! -\! x_j\! + \! 2\eta )
\sigma (x_i\! -\! x_j-\eta)}{\sigma (x_i-x_j \! +\! \eta )
\sigma (x_i \! -\! x_j)}-
\prod_{j\neq i}
\frac{\sigma (x_i\! -\! x_j\! -\!  2\eta )
\sigma (x_i\! -\! x_j+\eta )}{\sigma (x_i\! -\! x_j\! -\! \eta )
\sigma (x_i\! -\! x_j)}\right ]}=0.
\end{array}
\eeq
As is shown below, the properly taken limit $\eta \to 0$ 
of this equation coincides with
equation (\ref{bkp7}). In the rational limit one should substitute 
$\zeta (x)\to 1/x$, $\sigma (x)\to x$. 

In Appendix C we show that the same equations can be
obtained by restriction of the Ruijsenaars-Schneider dynamics 
with respect
to the time flow $\p_{t_1}-\p_{\bar t_1}$ of the system containing $2N$ particles to the half-dimensional 
subspace of the $4N$-dimensional phase space corresponding
to the configuration in which the particles stick together  
in pairs such that the distance between particles in each pair is equal to
$\eta$. This configuration is destroyed by the flow 
$\p_{t_1}+\p_{\bar t_1}$ but is preserved by the flow
$\p_{t_1}-\p_{\bar t_1}$. We conjecture that
it is preserved also by all higher flows 
$\p_{t_k}-\p_{\bar t_k}$. The time evolution in $t=t_1-\bar t_1$  
of the pairs with coordinates $x_i$ is given by equations (\ref{b13}).

\subsubsection{The limit $\eta \to 0$}

The ``non-relativistic limit'' $\eta \to 0$ in (\ref{b9}) yields
\beq\label{lim1}
v(x)=1+\eta ^2 u(x)  + O(\eta^4),
\eeq
where 
$u(x)$ is given by (\ref{bkp6}). Then the limit of the difference
operator $v(x)(e^{\eta \p_x}-e^{-\eta \p_x})$ is
\beq\label{lim2}
v(x)(e^{\eta \p_x}-e^{-\eta \p_x})=2\eta \p_x +
\frac{\eta^3}{3}\Bigl (\p_x^3 +6u(x)\p_x \Bigr ) +O(\eta^5),
\eeq
i.e., in the next-to-leading order the differential operator
$\p_x^3 +6u\p_x$ participating in the linear problem for the BKP equation
arises (see equation (\ref{bkp1})). Let us pass to the variables
\beq\label{lim3}
X=x+2\eta t, \quad T=\frac{\eta^3}{3}\, t,
\eeq
then $\p_X =\p_x$, $\p_T = \displaystyle{\frac{3}{\eta^3}\, 
(\p_t -2\eta \p_x)}$ and in the limit $\eta \to 0$ 
the linear problem (\ref{b1}) becomes $\p_T \psi =(\p_X^3 -6u\p_X)\psi$
which is the linear problem (\ref{bkp1}) for the BKP equation. 

Taking the change of variables (\ref{lim3}) into account, let us 
find the $\eta \to 0$ limit of equation (\ref{b13}). We have:
$$
\sigma (x-x_i)=\sigma \Bigl (X-\frac{6}{\eta^2}\, T -x_i \Bigr )=
\sigma (X-X_i),
$$
whence $X_i= \displaystyle{\frac{6}{\eta^2}T +x_i}$ and
$\displaystyle{
\dot x_i =\p_t x_i =-2\eta + \frac{\eta^3}{3}\, \p_T \! X_i}
$. Expanding equation (\ref{b13}) in powers of $\eta$, we obtain:
$$
\frac{\eta^6}{9}\, \p_T^2 \! X_i -4\eta^2 \sum_{k\neq i}
\Bigl (1\! -\! \frac{\eta^2}{6}\, \p_T \! X_i\Bigr )
\Bigl (1\! -\! \frac{\eta^2}{6}\, \p_T \! X_k\Bigr )
\Bigr ( \eta^2 \wp '(X_{ik})\! +\! \frac{\eta^4}{12}\, \wp '''(X_{ik})
\! +\! O(\eta^6)\Bigr )
\! -\! U_i=0,
$$
where $X_{ik}\equiv X_i-X_k$ and
$$
U_i=\sigma (2\eta )
\left [
\prod_{j\neq i}\frac{\sigma (X_{ij}+2\eta )\sigma (X_{ij}-\eta )}{\sigma
(X_{ij}+\eta )\sigma (X_{ij})}-
\prod_{j\neq i}\frac{\sigma (X_{ij}-2\eta )\sigma (X_{ij}+\eta )}{\sigma
(X_{ij}-\eta )\sigma (X_{ij})}\right ]
$$
$$
=-4\eta^4 \sum_{j\neq i}\wp '(X_{ij})+8\eta^6 \sum_{j\neq i}\wp (X_{ij})
\sum_{l\neq i}\wp '(X_{il}) -\eta^6 \sum_{j\neq i}\wp '''(X_{ij})
+O(\eta^8).
$$
It is easy to see that the terms of order $\eta^4$ cancel in the equation
and in the leading order $\eta^6$ equation (\ref{bkp7}) arises.

\subsubsection{Existence of meromorphic wave solutions}

Similarly to the previous cases, equation
(\ref{b8}) is a sufficient condition for local existence of a meromorphic
wave solution to equation (\ref{b1}) depending on a spectral parameter $k$
with a pole at $x=a$ and regular 
at the points $x=a\pm \eta$. However, the proof requires more 
sophisticated calculations than in the Toda lattice case.
To proceed, we represent $v(x)$ in the form 
\beq\label{b14}
v(x)=\frac{\tau (x+\eta )\tau (x-\eta )}{\tau^2(x)}
\eeq
which is motivated by the result of \cite{KZ22} ($\tau (x)$ is 
the tau-function of the Toda lattice of type B). We assume that
$\tau (x)$ has a simple zero at the point $a$ and that it is regular
and non-zero in some neighborhood of this point including the 
points $a\pm \eta$:
\beq\label{b15}
\tau (x)=(x-a)\rho (x-a),
\eeq
where the function $\rho (x)$ is regular and non-zero at $x=0$.
It depends also on the time $t$.
Then the coefficients in (\ref{b2}), (\ref{b3}) are expressed as
\beq\label{b16}
\nu =-\eta^2 \frac{\rho (\eta )\rho (-\eta )}{\rho^2(0)},
\quad
\mu =\nu \left (\frac{\rho '(\eta )}{\rho (\eta )}+
\frac{\rho '(-\eta )}{\rho (-\eta )}-2
\frac{\rho '(0)}{\rho (0)}\right ),
\eeq
\beq\label{b16a}
V^{\pm}(a)=\pm \frac{2}{\eta}\, \frac{\rho (0)\rho 
(\pm 2\eta )}{\rho^2(\pm \eta )}.
\eeq
It is also convenient to introduce the function
\beq\label{b17}
\varphi_+(x)=\log \frac{\tau (x-\eta )}{\tau (x)}.
\eeq
The function $\dot \varphi_+ (x)$ has simple poles at the points 
$x=a$ and $x=a+\eta$ with the expansions
\beq\label{b18}
\dot \varphi_+ (x)=\left \{ \begin{array}{l}
\displaystyle{ \frac{\dot a}{x-a} +\varphi_0 +\ldots \, , \quad x\to a,}
\\ \\
\displaystyle{-\frac{\dot a}{x\! -\! a\! -\! \eta} +\varphi_1 +
\ldots \, , \quad x\to a+\eta ,}
\end{array}
\right.
\eeq
where
\beq\label{b19}
\begin{array}{l}
\displaystyle{
\varphi_0 = \dot a \left (\frac{1}{\eta}-
\frac{\rho '(-\eta )}{\rho (-\eta )}+\frac{\rho '(0)}{\rho (0)}\right )
+\frac{\dot \rho (-\eta )}{\rho (-\eta )}-\frac{\dot \rho (0)}{\rho (0)}},
\\ \\
\displaystyle{
\varphi_1 = \dot a \left (\frac{1}{\eta}\, +\, 
\frac{\rho '(\eta )}{\rho (\eta )}\, -\, 
\frac{\rho '(0)}{\rho (0)}\right )
-\frac{\dot \rho (\eta )}{\rho (\eta )}+\frac{\dot \rho (0)}{\rho (0)}}.
\end{array}
\eeq
It is easy to check that
$$
\varphi_0-\varphi_1=\frac{\dot \nu}{\nu}-\frac{\mu}{\nu}\, \dot a.
$$

\begin{proposition}
Assume that $v(x)$
in (\ref{b1}) is represented in the form
(\ref{b14}), $\tau (x)$ has a simple zero at a point $a$ and
$\tau (a+\eta )\tau (a-\eta )\neq 0$. 
If
condition (\ref{b8}) holds, 
then all wave solutions of equation (\ref{b1})
of the form
\beq\label{b20}
\psi (x)= k^{x/\eta}e^{kt+\varphi_+(x)} 
\Bigl (1+ \sum_{s\geq 1} \xi_s(x)k^{-s}\Bigr ),
\quad k\to \infty ,
\eeq
where $\varphi_+(x)$ is given by (\ref{b17}) are meromorphic
in a neighborhood of the point $a$
with a simple pole at $x=a$ and regular at
$x=a\pm \eta$.
\end{proposition}

\noindent
{\it Proof.}
Substituting the series (\ref{b20}) into the equation (\ref{b1}), 
we obtain the recurrence relation
for the coefficients $\xi_s$:
\beq\label{b21}
\xi_{s+1}(x+\eta )-\xi_{s+1}(x)=
\dot \varphi_+ (x)\xi_s(x)+\dot \xi_s (x)
+v(x)v(x-\eta )\xi_{s-1}(x-\eta ), \quad s\geq 0,
\eeq
where we set $\xi_{-1}=0$, $\xi_0=1$. 
The factor $e^{\varphi_+ (x)}$ in (\ref{b20}) has a pole at $x=a$ and
a zero at $x=a+\eta$. This means that the coefficients $\xi_s(x)$ 
are regular at $x=a$ and may have a pole at $x=a+\eta$.
Let the Laurent expansion of $\xi_s$ near the point $a+\eta$ be
\beq\label{b22}
\xi_s(x)=\frac{r_s}{x-a-\eta} +r_{s,0}+O(x-a-\eta )\, , 
\quad x\to a+\eta .
\eeq
The meromorphic wave solution with the required properties exists
if the sum of residues of the right hand side of (\ref{b21}) at the 
points $x=a$ and $x=a+\eta$ is equal to zero for all $s\geq 0$. 
This can be proved by induction in a similar way as before
but the calculations are more involved. 

To proceed, we need some properties of the coefficient function
$v(x)v(x-\eta )$ in (\ref{b21}). We have:
$$
v(x)v(x-\eta )=\frac{\tau (x-2\eta )\tau (x+\eta )}{\tau (x-\eta )
\tau (x)},
$$
whence
\beq\label{b23}
v(x)v(x-\eta )=\frac{\nu V^+(a)}{x-a-\eta}+O(1), \quad x\to a+\eta ,
\qquad
v(x)v(x-\eta )\Bigr |_{x=a-\eta}=0,
\eeq
\beq\label{b24}
v(x)v(x-\eta )=\frac{\nu V^-(a)}{x-a}+\Omega \nu V^-(a)+O(x-a), \quad
x\to a,
\eeq
where
\beq\label{b25}
\Omega = \frac{3}{2\eta}+\frac{\rho '(-2\eta )}{\rho (-2\eta )}-
\frac{\rho '(-\eta )}{\rho (-\eta )}+\frac{\rho '(\eta )}{\rho (\eta )}
-\frac{\rho '(0)}{\rho (0)}.
\eeq

Expanding equation (\ref{b21}) near the point $x=a+\eta$, we get, equating
the coefficients in front of the poles:
\beq\label{b26}
r_{s+1}=\dot a r_{s,0} -\varphi_1 r_s -\dot r_s -\nu V^+(a) \xi_{s-1}(a).
\eeq
Similarly, expanding equation (\ref{b21}) 
near the point $x=a$, we get, equating
the coefficients in front of $(x-a)^{-1}$ and $(x-a)^0$:
\beq\label{b27}
\left \{ \begin{array}{l}
r_{s+1}=\dot a \xi_s(a)+\nu V^-(a)\xi_{s-1}(a-\eta ),
\\ \\
r_{s+1, 0}-\xi_{s+1}(a)= \dot \xi_s(a) +
\varphi_0 \xi_s(a) +\nu V^-(a)\xi_{s-1}'(a-\eta )+
\nu V^-(a)\Omega \xi_{s-1}(a-\eta ),
\end{array}
\right.
\eeq
where 
$
\dot \xi_s(a)=
\dot a \xi_s'(a)+\p_t \xi_s(a)
$
is the full time derivative. At the point $x=a-\eta$ 
all terms in equation (\ref{b21}) are regular and the equation
gives:
\beq\label{b28}
\xi_{s+1}(a)-\xi_{s+1}(a-\eta )=\p_t \varphi_+(a-\eta )\xi_s (a-\eta )
+\p_t \xi_s (a-\eta )
\eeq
(the last term in (\ref{b21}) vanishes at this point). 

Let 
$$
R_s=\dot a (r_{s,0}-\xi_s(a))-\varphi_1
r_s -\dot r_s -\nu V^+(a)\xi_{s-1}(a)-\nu V^-(a)\xi_{s-1}(a-\eta )
$$
be sum of the residues at the left hand side of 
(\ref{b21}) at the points $x=a$ and $x=a+\eta$ which 
must be zero. It is seen from (\ref{b21}) that $R_1=0$. Our induction
assumption is that $R_s=0$ for some $s$; let us show that this implies
that $R_{s+1}=0$. We have:
$$
R_{s+1}=\dot a (r_{s+1,0}-\xi_{s+1}(a))-\varphi_1
r_{s+1} -\dot r_{s+1} -\nu V^+(a)\xi_{s}(a)-\nu V^-(a)\xi_{s}(a-\eta ).
$$
A straightforward calculation which uses recurrence relations
(\ref{b26}), (\ref{b27}) and (\ref{b28}) yields:
$$
R_{s+1}=\left [ \dot a\Bigl (\frac{\dot \nu}{\nu}-\frac{\mu}{\nu}\, 
\dot a \Bigr )-\ddot a -\nu (V^+(a)+V^-(a))\right ]-\p_t R_s
$$
$$
+\nu V^-(a)\xi_{s-1}(a-\eta )\left [ \dot a\Omega -\varphi_1 -
\p_t \log (\nu V^-(a)) +\p_t \varphi_+(a-\eta )\right ].
$$
The first two terms vanish by virtue of the condition (\ref{b8}) and
the induction assumption. Using equations (\ref{b16}), (\ref{b16a}),
(\ref{b19}) and (\ref{b25}), one can show that the third term also vanishes. 
Therefore, we have proved that from $R_s=0$ it follows that 
$R_{s+1}=0$ which implies the existence of a meromorphic wave solution. 
\square

\section{Fully difference equation}

The case of fully difference equation was considered by one of the
authors in \cite{K10} but we find it appropriate to include it here
for completeness. 

Let us consider the difference equation
\beq\label{d1}
\psi_{t+1}(x)=\psi_t(x+\eta )+ u_t(x)\psi_t (x), 
\qquad 
u_t(x)=\frac{\tau_t(x) \tau_{t+1}(x+\eta )}{\tau_t(x+\eta )\tau_{t+1}(x)}
\eeq
which serves as the auxiliary linear problem for the Hirota bilinear
difference equation for the tau-function $\tau_t(x)$ \cite{Hirota,Miwa}.

\begin{proposition}
Let $\tau_t(x)$ have a simple zero at some point $a_t$:
$\tau_t(a_t)=0$, $\tau_t '(a_t)\neq 0$ and $\tau_t(a_t-\eta )
\tau_{t+1}(a_t)\neq 0$. Then the necessary condition that equation
(\ref{d1}) has a meromorphic solution with a simple pole
at $x=a_t$ 
regular at $x=a_t \pm \eta$, $x=a_{t+1}$
is
\beq\label{d4}
\frac{\tau_{t+1}(a_t)\tau_t(a_t-\eta )
\tau_{t-1}(a_t+\eta )}{\tau_{t+1}(a_t-\eta )\tau_t(a_t+\eta )
\tau_{t-1}(a_t)}=-1.
\eeq
\end{proposition}

\noindent
{\it Proof.}
Tending $x$ to $a_{t+1}$, $a_t-\eta$ and $a_{t+1}-\eta$ in equation
(\ref{d1}), we obtain the relations
\beq\label{d3}
\left \{ \begin{array}{l}
\displaystyle{
\alpha_{t+1}=\frac{\tau_t(a_{t+1})
\tau_{t+1}(a_{t+1}+\eta )}{\tau_{t+1}'(a_{t+1})\tau_t(a_{t+1}+\eta )}\,
\psi_t(a_{t+1}),}
\\ \\
\displaystyle{
\alpha_{t}=-\frac{\tau_t(a_{t}-\eta )
\tau_{t+1}(a_{t})}{\tau_{t}'(a_{t})\tau_{t+1}(a_{t}-\eta )}\,
\psi_t(a_{t}-\eta ),}
\\ \\
\psi_{t+1}(a_{t+1}-\eta )=\psi_t(a_{t+1}).
\end{array}
\right.
\eeq
Combining them, we arrive at the condition (\ref{d4}).
\square

Let $\tau_t(x)$ be an ``elliptic polynomial'' of the form
\beq\label{d5}
\tau_t(x)=\prod_{j=1}^N \sigma (x-x_j^t)
\eeq
with $N$ simple roots $x_j^t$ in the fundamental domain,
then the conditions (\ref{d4}) with $a_t=x_i^t$ for each 
$i=1, \ldots , N$ yield the equations
\beq\label{d6}
\prod_{j=1}^N \frac{\sigma (x_i^t-x_j^{t+1})\sigma (x_i^t -x_j^t -\eta )
\sigma (x_i^t-x_j^{t-1}+\eta )}{\sigma (x_i^t-x_j^{t+1}-\eta )
\sigma (x_i^t -x_j^t +\eta )
\sigma (x_i^t-x_j^{t-1})}=-1.
\eeq
These are equations of motion for the discrete time version of the 
Ruijsenaars-Schneider system first obtained in \cite{NRK96}.

Finally, we will show that (\ref{d4}) is a sufficient condition 
for local existence of a meromorphic wave solution 
to equation (\ref{d1}) of the form
\beq\label{d7}
\psi_t(x)=k^{x/\eta}k^t \Bigl (1+\sum_{s\geq 1}\xi_s^t(x)k^{-s}\Bigr )
\eeq
having a simple pole at $x=a_t$ and regular at $x=a_t\pm \eta$, $x=a_{t+1}$.

\begin{proposition}
Assume that $\tau_t (x)$ in (\ref{d1}) 
has a zero at some point $a_t$
such that $\tau_t '(a_t)\neq 0$, $\tau_t(a_t-\eta )
\tau_{t+1}(a_t)\neq 0$ and condition (\ref{d4}) holds. 
Then all wave solutions to equation (\ref{d1})
of the form
(\ref{d7}) are meromorphic
in a neighborhood of the point $a$
with a simple pole at $x=a$ and regular at
$x=a\pm \eta$.
\end{proposition}

\noindent
{\it Proof.}
Substituting the series (\ref{d7})
into the equation, we obtain the recurrence relation
\beq\label{d8}
\xi_{s+1}^{t+1}(x)-\xi_{s+1}^t(x+\eta )=u_t(x)\xi_s^t(x).
\eeq
Let the expansion of the function $\psi_t(x)$ near the pole be of the form
\beq\label{d9}
\psi_t(x)=\frac{r_s^t}{x-a_t}+r_{s,0}^t +O(x-a_t).
\eeq
Tending $x$ to $a_{t+1}$, $a_t-\eta$ and $a_{t+1}-\eta$ in equation
(\ref{d8}), we obtain the equalities
\beq\label{d10}
\left \{ \begin{array}{l}
\displaystyle{
r_{s+1}^{t+1}=\frac{\tau_t(a_{t+1})
\tau_{t+1}(a_{t+1}+\eta )}{\tau_{t+1}'(a_{t+1})\tau_t(a_{t+1}+\eta )}\,
\xi_s^t(a_{t+1}),}
\\ \\
\displaystyle{
r_{s+1}^t=-\frac{\tau_t(a_{t}-\eta )
\tau_{t+1}(a_{t})}{\tau_{t}'(a_{t})\tau_{t+1}(a_{t}-\eta )}\,
\xi_s^t(a_{t}-\eta ),}
\\ \\
\xi_{s}^{t+1}(a_{t+1}-\eta )=\xi_{s}^t(a_{t+1}),
\end{array}
\right.
\eeq
where we shifted $s\to s-1$ in the last equation. Combining them and taking
into account the condition (\ref{d4}), we see that the two expressions 
for $r_{s+1}^t$ that follow from (\ref{d8}) actually coincide 
whence we conclude that the solution of the form (\ref{d7}) exists. 
\square

\section{Concluding remarks}

In this paper we have further developed the approach to integrable 
many-body systems based on monodromy free linear equations, 
i.e. on finding
conditions of existence of meromorphic
solutions to linear partial 
differential and difference equations for all cases 
which arise as linear problems for integrable 
nonlinear equations. Some of them were previously discussed by 
one of the authors in \cite{K05,K10,K22}. We have also shown that 
these conditions are simultaneously sufficient conditions for local 
existence of meromorphic wave solutions depending on a spectral parameter.
These conditions straightforwardly lead to equations of motion for poles
of elliptic solutions to nonlinear integrable equations. It turns out
that the dynamics 
of poles is isomorphic to many-body systems of Calogero-Moser type
which include the Calogero-Moser system itself, its relativistic
extension (the Ruijsenaars-Schneider system) and a rather
exotic system with
three-body interaction found in \cite{RZ20}. We note that the approach
of this paper is the shortest way to obtain the equations of motion. 
Presumably, all these systems are integrable as they arise as certain
finite-dimensional reductions of integrable nonlinear systems with
infinitely many degrees of freedom. Independent proofs of integrability
exist for the Calogero-Moser and Ruijsenaars-Schneider systems while 
for the system introduced in \cite{RZ20} this is still a hypothesis. 

The main new result of this paper is the many-body system 
with equations of motion 
(\ref{int1}), (\ref{int2}) representing dynamics of 
poles of elliptic solutions to the Toda lattice of type B recently
introduced in \cite{KZ22}. This system can be regarded as a kind
of relativistic extension of the system found in \cite{RZ20}
with equations of motion (\ref{bkp7})
in the sense that the former is related to the latter in the same way
as the Ruijsenaars-Schneider system is related to the Calogero-Moser
system. 

There are some interesting open problems. First, it is
not clear whether the system (\ref{int1}), (\ref{int2}) is Hamiltonian.
Second, a commutation representation for it is not yet known.
Presumably, such commutation representation is of the form of
the Manakov's triple as it is the case for the system 
(\ref{bkp7}) which arises from (\ref{int1}), (\ref{int2})
in the $\eta \to 0$ limit. Last but not least, there is the problem 
of proving integrability of the system (\ref{int1}), (\ref{int2}).
At the moment we know only one integral of motion which is
$\displaystyle{I=\sum_i \dot x_i}$. Its conservation 
($\ddot I=0$) can be seen directly 
from the equations of motion taking into account
that
$$
\sum_{i=1}^N U(x_{i1}, \ldots , x_{iN})=0
$$
which follows from the fact that the left hand side is sum of the 
residues at the $2N$ poles of the elliptic function
$$
f(x)=\prod_{j=1}^N \frac{\sigma (x-x_j+2\eta )
\sigma (x-x_j-\eta )}{\sigma (x-x_j+\eta )\sigma (x-x_j)}
$$
at the points $x=x_i$, $x=x_i-\eta$, $i=1, \ldots \, , N$.

\section*{Appendix A: The Weierstrass functions}
\addcontentsline{toc}{section}{Appendix A: The Weierstrass functions}
\def\theequation{A\arabic{equation}}
\def\theHequation{\theequation}
\setcounter{equation}{0}

Throughout the main text, we use the standard Weierstrass functions:
the $\sigma$-function, the $\zeta$-function and the $\wp$-function. 

The Weierstrass $\sigma$-function 
with quasi-periods $2\omega$, $2\omega '$ such that 
${\rm Im} (\omega '/ \omega )>0$ is defined by the infinite product
\beq\label{A1}
\sigma (x)=\sigma (x |\, \omega , \omega ')=
x\prod_{s\neq 0}\Bigl (1-\frac{x}{s}\Bigr )\, 
e^{\frac{x}{s}+\frac{x^2}{2s^2}},
\quad s=2\omega m+2\omega ' m' \quad \mbox{with integer $m, m'$}.
\eeq 
It is an odd entire quasiperiodic function in the complex plane. 
The Weierstrass $\zeta$-function is defined as
\beq\label{A2}
\zeta (x)=\frac{\sigma '(x)}{\sigma (x)}=
\frac{1}{x}+\sum_{s\neq 0} \Bigl ( \frac{1}{x-s}+\frac{1}{s}+
\frac{x}{s^2}\Bigr ).
\eeq
It is an odd function with first order poles at the points 
of the lattice $s=2\omega m+2\omega ' m'$ with integer $m, m'$.
The definition of the Weierstrass $\wp$-function is 
\beq\label{A3}
\wp (x)=-\zeta '(x) = \frac{1}{x^2}+\sum_{s\neq 0} \Bigl ( 
\frac{1}{(x-s)^2}-\frac{1}{s^2}\Bigr ).
\eeq
It is an even double-periodic function with periods $2\omega , 2\omega '$
and with second order poles at the points 
of the lattice $s=2\omega m+2\omega ' m'$ with integer $m, m'$.

The monodromy properties of the $\sigma$-function 
under shifts by the quasi-periods
are as follows:
\beq\label{A4}
\begin{array}{l}
\sigma (x+2\omega )=-e^{2\zeta (\omega )(x+\omega )}\sigma (x),
\\ \\
\sigma (x+2\omega ' )=-e^{2\zeta (\omega ')(x+\omega ' )}\sigma (x).
\end{array}
\eeq
The $\zeta$-function acquires an additive constant when the argument
is shifted by any quasi-period:
\beq\label{A5}
\begin{array}{l}
\zeta (x+2\omega )=\zeta (x)+\zeta (\omega ),
\\ \\
\zeta (x+2\omega ' )=\zeta (x)+\zeta (\omega ').
\end{array}
\eeq
These constants are related by the identity
$2\omega ' \zeta (\omega )-2\omega \zeta (\omega ')=\pi i$.

\section*{Appendix B: The Ruijsenaars-Schneider model}
\addcontentsline{toc}{section}{Appendix B: The Ruijsenaars-Schneider model}
\def\theequation{B\arabic{equation}}
\def\theHequation{\theequation}
\setcounter{equation}{0}

Here we collect some facts on the elliptic Ruijsenaars-Schneider system \cite{RS86}
following the paper \cite{Ruij87}.
The $N$-particle elliptic Ruijsenaars-Schneider system
is a completely integrable model.
The canonical Poissson brackets between coordinates and momenta are
$\{x_i, p_j\}=\delta_{ij}$.
The integrals of motion in involution have the form
\beq\label{intr1}
I_k= \sum_{I\subset \{1, \ldots , N\}, \, |I|=k}
\exp \Bigl (\sum_{i\in I}p_i\Bigr ) \prod_{i\in I, j\notin I}\frac{\sigma
(x_i-x_j+\eta )}{\sigma (x_i-x_j)}, \quad k=1, \ldots , N.
\eeq
It is convenient to put $I_0=1$.
Important particular cases of (\ref{intr1}) are
\beq\label{intr2}
I_1= H_1=\sum_i e^{p_i}\prod_{j\neq i} \frac{\sigma
(x_i-x_j+\eta )}{\sigma (x_i-x_j)}
\eeq
which is the Hamiltonian $H_1$ of the chiral Ruijsenaars-Schneider model and
\beq\label{intr2a}
I_N=\exp \Bigl (\sum_{i=1}^{N}p_i\Bigr ).
\eeq

Let us denote the time variable of the Hamiltonian flow with the Hamiltonian $H_1$
by $t_1$.
The velocities of the particles are
\beq\label{intr4}
\dot x_i =\frac{\p H_1}{\p p_i}=e^{p_i}\prod_{j\neq i} \frac{\sigma
(x_i-x_j+\eta )}{\sigma (x_i-x_j)},
\eeq
where dot means the $t_1$-derivative.
The Hamiltonian equations $\dot x_i=\p H_1/\p p_i$, 
$\dot p_i=-\p H_1/ \p x_i$ are equivalent to the following
equations of motion:
\beq\label{te4}
\begin{array}{lll}
\ddot x_i &=&\displaystyle{-\sum_{k\neq i}\dot x_i\dot x_k \Bigl (
\zeta (x_i-x_k+\eta )+\zeta (x_i-x_k-\eta )-2\zeta (x_i-x_k)\Bigr )}
\\ && \\
&=&\displaystyle{\sum_{k\neq i}\dot x_i\dot x_k
\frac{\wp '(x_i-x_k)}{\wp (\eta )-
\wp (x_i-x_k)}.}
\end{array}
\eeq
The properly taken limit $\eta \to 0$ (the ``non-relativistic limit'')
gives equations of motion of the elliptic Calogero-Moser system.

One can also introduce integrals of motion $I_{-k}$ as
\beq\label{intr1b}
I_{-k}=I_{N}^{-1}I_{N-k}=\sum_{I\subset \{1, \ldots , N\}, \, |I|=k}
\exp \Bigl (-\sum_{i\in I}p_i\Bigr ) \prod_{i\in I, j\notin I}\frac{\sigma
(x_i-x_j-\eta )}{\sigma (x_i-x_j)}.
\eeq
In particular,
\beq\label{intr2b}
I_{-1}= \sum_i e^{-p_i}\prod_{j\neq i} \frac{\sigma
(x_i-x_j-\eta )}{\sigma (x_i-x_j)}.
\eeq
It can be easily verified that equations of motion
in the time $\bar t_1$ corresponding to the Hamiltonian 
$\bar H_1=\sigma^2(\eta )I_{-1}$
are the same
as (\ref{te4}).

The ``physical'' Hamiltonian of the Ruijsenaars-Schneider model
is $H_+=H_1 +\bar H_1$. Below in Appendix C we consider the Hamiltonian
flow corresponding to the Hamiltonian $H_-=H_1 -\bar H_1$.

\section*{Appendix C: How Ruijsenaars-Schneider particles \\ 
stick together}
\addcontentsline{toc}{section}{Appendix C: 
How Ruijsenaars-Schneider particles stick together}
\def\theequation{C\arabic{equation}}
\def\theHequation{\theequation}
\setcounter{equation}{0}

In this appendix we show how to restrict the Ruijsenaars-Schneider
dynamics of the $N=2n$-particle system to the subspace in which the
particles stick together in $n$ pairs such that
\beq\label{C1}
x_{2i}-x_{2i-1}=\eta , \qquad i=1, \ldots , n.
\eeq
We introduce the variables 
\beq\label{C2}
X_i =x_{2i-1}, \qquad i=1, \ldots , n
\eeq
which are coordinates of the pairs. Such configuration is destroyed 
by the $H_+$-Ha\-mil\-to\-ni\-an flow 
$\p_{t_1}+\p_{\bar t_1}$ but is preserved
by the $H_-$-flow $\p_t=\p_{t_1}-\p_{\bar t_1}$, as we shall see below.

The Hamiltonian $H_-$ reads
$$
H_-=\sum_i e^{p_i}\prod_{j\neq i} \frac{\sigma
(x_i-x_j+\eta )}{\sigma (x_i-x_j)}-\sigma^2(\eta )
\sum_i e^{-p_i}\prod_{j\neq i} \frac{\sigma
(x_i-x_j-\eta )}{\sigma (x_i-x_j)}.
$$
For the velocities $\dot x_i =\p H_-/\p p_i$ we have:
$$
\dot x_{2i-1}=e^{p_{2i-1}}\! \prod_{j=1, \neq 2i-1}^{2n} \! \frac{\sigma
(x_{2i-1, j}+\eta )}{\sigma (x_{2i-1, j})}+
\sigma^2(\eta )e^{-p_{2i-1}}\! \prod_{j=1, \neq 2i-1}^{2n} \! \frac{\sigma
(x_{2i-1, j}-\eta )}{\sigma (x_{2i-1, j})},
$$
$$
\dot x_{2i}=e^{p_{2i}}\! \prod_{j=1, \neq 2i}^{2n} \! \frac{\sigma
(x_{2i, j}+\eta )}{\sigma (x_{2i, j})}\, +\, 
\sigma^2(\eta )e^{-p_{2i}}\! \prod_{j=1, \neq 2i}^{2n} \! \frac{\sigma
(x_{2i, j}-\eta )}{\sigma (x_{2i, j})},
$$
where $x_{ik}\equiv x_i-x_k$. Suppose that 
the momenta remain finite under the
restriction to (\ref{C1}). Then in terms of coordinates $X_i$ of the pairs  
we have from these formulas:
\beq\label{C3}
\begin{array}{l}
\displaystyle{
\dot x_{2i-1}=\sigma (\eta )\sigma (2\eta )e^{-p_{2i-1}}
\! \prod_{j=1, \neq i}^n \! \frac{\sigma (X_{ij}-2\eta )}{\sigma (X_{ij})}},
\\ \\
\displaystyle{
\dot x_{2i}=\frac{\sigma (2\eta )}{\sigma (\eta )}
\, e^{p_{2i}}
\! \prod_{j=1, \neq i}^n \! \frac{\sigma (X_{ij}+2\eta )}{\sigma (X_{ij})}}.
\end{array}
\eeq
The dynamics preserves the configuration (\ref{C1}) if $\dot x_{2i-1}=
\dot x_{2i}$, whence we should require
$$
e^{p_{2i-1}+p_{2i}}=\sigma^2(\eta )\prod_{j\neq i}
\frac{\sigma (X_{ij}-2\eta )}{\sigma (X_{ij}+2\eta )}.
$$
Resolving this constraint, we can put
\beq\label{C4}
p_{2i-1}=\alpha_i +P_i, \quad p_{2i}=\alpha_i -P_i, \quad i=1, \ldots , n,
\eeq
where
\beq\label{C5}
\alpha_i= \log \sigma (\eta )+\frac{1}{2}\sum_{j\neq i}
\log \frac{\sigma (X_{ij}-2\eta )}{\sigma (X_{ij}+2\eta )}
\eeq
and $P_i$ are arbitrary. We have thus restricted the original 
$4n$-dimensional phase space ${\cal F}$ with coordinates 
$\{p_j, x_j\}$, $j=1, \ldots , 2n$ to 
the $2n$-dimensional subspace ${\cal P}$ 
with coordinates $\{P_i, X_i\}$,
$i=1, \ldots , n$ corresponding to joining the particles into pairs
of the form (\ref{C1}). Equations (\ref{C3}) are then equivalent to
\beq\label{C6}
\dot X_i=\sigma (2\eta )e^{-P_i}\prod_{j\neq i}
\frac{(\sigma (X_{ij}-2\eta )
\sigma (X_{ij}+2\eta ))^{1/2}}{\sigma (X_{ij})}.
\eeq

Let us now pass to the second set of the Hamiltonian equations, 
$\dot p_i=-\p H_-/\p x_i$:
\beq\label{C7a}
\begin{array}{lll}
\dot p_i & =& 
\displaystyle{
\sigma^2(\eta )e^{-p_i}\! \prod_{k=1, \neq i}^{2n}
\frac{\sigma (x_{ik}-\eta )}{\sigma (x_{ik})}\sum_{j=1, \neq i}^{2n}
\Bigl (\zeta (x_{ij}-\eta )-\zeta (x_{ij})\Bigr )}
\\ && \\
&&\displaystyle{
-\, e^{p_i}\! \prod_{k=1, \neq i}^{2n}
\frac{\sigma (x_{ik}+\eta )}{\sigma (x_{ik})}\sum_{j=1, \neq i}^{2n}
\Bigl (\zeta (x_{ij}+\eta )-\zeta (x_{ij})\Bigr )}
\\ &&\\
&&\displaystyle{
+\, \sigma^2(\eta )\sum_{l=1, \neq i}^{2n}\!
e^{-p_l}\! \prod_{k=1, \neq l}^{2n}
\frac{\sigma (x_{lk}-\eta )}{\sigma (x_{lk})}
\Bigl (\zeta (x_{il}+\eta )-\zeta (x_{il})\Bigr )}
\\ &&\\
&&\displaystyle{
-\, \sum_{l=1, \neq i}^{2n}\!
e^{p_l}\! \prod_{k=1, \neq l}^{2n}
\frac{\sigma (x_{lk}+\eta )}{\sigma (x_{lk})}
\Bigl (\zeta (x_{il}-\eta )-\zeta (x_{il})\Bigr )}.
\end{array}
\eeq
Restricting to the subspace ${\cal P}$, we have:
\beq\label{C7}
\begin{array}{lll}
\dot p_{2i-1} & \!\! =\!\! & 
\displaystyle{
\sigma (\eta )\sigma (2\eta )e^{-\alpha_i -P_i}\!\! \prod_{k=1, \neq i}^n
\!\!
\frac{\sigma (X_{ik}-2\eta )}{\sigma (X_{ik})}\left [
\sum_{j=1, \neq i}^n\Bigl 
(\zeta (X_{ij}\! -\! 2\eta )\! -\! \zeta (X_{ij})\Bigr )
\! +\! \zeta (\eta ) \! -\! \zeta (2\eta )\right ]}
\\ &&\\
&& +\, \displaystyle{
\sigma (\eta )\sigma (2\eta )\sum_{l=1, \neq i}^n
e^{-\alpha_l -P_l}\! \prod_{k=1, \neq l}^n
\frac{\sigma (X_{lk}-2\eta )}{\sigma (X_{lk})}
\Bigl (\zeta (X_{il}+\eta )-\zeta (X_{il})\Bigr )}
\\ &&\\
&& -\, \displaystyle{
\frac{\sigma (2\eta )}{\sigma (\eta )}\sum_{l=1}^n
e^{\alpha_l -P_l}\! \prod_{k=1, \neq l}^n
\frac{\sigma (X_{lk}+2\eta )}{\sigma (X_{lk})}
\Bigl (\zeta (X_{il}-2\eta )-\zeta (X_{il}-\eta )\Bigr )}
\\ && \\
&& +\, \displaystyle{
\sigma^{-1}(\eta )e^{\alpha_i +P_i}\prod_{k=1, \neq i}^n
\frac{\sigma (X_{ik}+\eta )}{\sigma (X_{ik})-\eta )}-
\sigma (\eta )e^{-\alpha_i +P_i}\prod_{k=1, \neq i}^n
\frac{\sigma (X_{ik}-\eta )}{\sigma (X_{ik})+\eta )}}.
\end{array}
\eeq
When passing from (\ref{C7a}) to (\ref{C7}) with the constraint 
(\ref{C1}), one encounters expressions like 
$\sigma (x_{2i}\! -\! x_{2i-1}\! -\! \eta )
\zeta (x_{2i}\! -\! x_{2i-1}\! -\! \eta )$
which is an indeterminacy of the form $0/0$. To resolve it, one
should put $x_{2i}\! -\! x_{2i-1}=\eta +\varepsilon$ and tend $\varepsilon
\to 0$.

Taking the time derivative of (\ref{C6}), we obtain:
$$
\ddot X_i=-\sigma (2\eta )\dot P_i e^{-P_i}
\prod_{j\neq i}
\frac{(\sigma (X_{ij}-2\eta )
\sigma (X_{ij}+2\eta ))^{1/2}}{\sigma (X_{ij})}
$$
$$
+\frac{1}{2}\sum_{j\neq i}\dot X_i (\dot X_i-\dot X_j)
\Bigl (\zeta (X_{ij}-2\eta )+\zeta (X_{ij}+2\eta )-2\zeta (X_{ij})\Bigr ),
$$
where we should substitute $\dot P_i=-\dot \alpha_i
+\dot p_{2i-1}$ from (\ref{C7}) taking into account
(\ref{C6}):
$$
\dot P_i =-\dot \alpha_i +\dot X_i \left [\sum_{j\neq i}
\Bigl (\zeta (X_{ij}-2\eta )-\zeta (X_{ij})\Bigr ) +\zeta (\eta )-
\zeta (2\eta )\right ]
$$
$$
+\sum_{l\neq i}\dot X_l \Bigl (\zeta (X_{il}+\eta )-\zeta (X_{il})\Bigr )
-\sum_{l}\dot X_l \Bigl (\zeta (X_{il}-2\eta )-\zeta (X_{il}-\eta )\Bigr )
$$
$$
+e^{P_i}\prod_{k\neq i}
\frac{\sigma^{1/2}(X_{ik}-2\eta )
\sigma (X_{ik}+\eta )}{\sigma^{1/2}(X_{ik}+2\eta )\sigma (X_{ik}-\eta )}
-e^{P_i}\prod_{k\neq i}\frac{\sigma^{1/2}(X_{ik}+2\eta )
\sigma (X_{ik}-\eta )}{\sigma^{1/2}(X_{ik}-2\eta )\sigma (X_{ik}+\eta )}.
$$
Substituting here
$$
\dot \alpha_i=\frac{1}{2} \sum_{j\neq i}(\dot X_i -\dot x_j)
\Bigl (\zeta (X_{ij}-2\eta )-\zeta (X_{ij}+2\eta )\Bigr ),
$$
we obtain, after cancellations:
\beq\label{C8}
\begin{array}{c}
\displaystyle{
\ddot X_i=-\sum_{j\neq i}\dot X_i \dot X_j \Bigl (
\zeta (X_{ij}+\eta )+\zeta (X_{ij}-\eta )-2\zeta (X_{ij})\Bigr )}
\\ \\
+\sigma (2\eta )
\displaystyle{
\left [
\prod_{j\neq i}\frac{\sigma (X_{ij}+2\eta )
\sigma (X_{ij}-\eta )}{\sigma
(X_{ij}+\eta )\sigma (X_{ij})}-
\prod_{j\neq i}\frac{\sigma (X_{ij}-2\eta )
\sigma (X_{ij}+\eta )}{\sigma (X_{ij}-\eta )\sigma (X_{ij})}\right ].
}
\end{array}
\eeq
These are equations (\ref{int1}), (\ref{int2}). 

A similar calculation for $\dot p_{2i}$ leads to the same result.
This means
that the restriction to the subspace ${\cal P}$ is consistent.

\section*{Acknowledgments}

\addcontentsline{toc}{section}{Acknowledgments}

The work of A.Z. (sections 2.2, 2.3, 3.3)
was supported by the 
Russian Science Foundation under grant 19-11-00062.


\begin{thebibliography}{99}

\addcontentsline{toc}{section}{References}

\bibitem{AMM77}
H. Airault, H.P. McKean, and J. Moser, {\it Rational and 
elliptic solutions of the
Korteweg-De Vries equation and a related many-body problem},
Commun. Pure Appl. Math., {\bf 30} (1977) 95--148.

\bibitem{Krichever78}
I.M. Krichever, {\it Rational solutions of the Kadomtsev-Petviashvili
equation and integrable systems of $N$ particles on a line},
Funct. Anal. Appl. {\bf 12:1} (1978) 59--61.

\bibitem{CC77} D.V. Chudnovsky, G.V. Chudnovsky, {\it Pole expansions of non-linear
partial differential equations}, Nuovo Cimento {\bf 40B} (1977) 339--350.

\bibitem{Calogero71}
F. Calogero, {\it Solution of the one-dimensional
$N$-body problems with quadratic
and/or inversely quadratic pair potentials}, J. Math. Phys.
{\bf 12} (1971) 419–-436.

\bibitem{Calogero75} F. Calogero, {\it Exactly solvable one-dimensional many-body
systems}, Lett. Nuovo Cimento {\bf 13} (1975) 411--415.

\bibitem{Moser75}
J. Moser, {\it Three integrable Hamiltonian systems connected with isospectral
deformations}, Adv. Math. {\bf 16} (1975) 197--220.

\bibitem{Krichever80} I.M. Krichever, {\it Elliptic 
solutions of the Kadomtsev-Petviashvili
equation and integrable systems of particles}, 
Funk. Anal. i Ego Pril. {\bf 14:4} (1980) 45--54
(in Russian); English translation:
Functional Analysis and Its Applications {\bf 14:4} (1980) 282–-290.

\bibitem{OP81} M.A. Olshanetsky and A.M. Perelomov, {\it Classical integrable
finite-dimensional systems related to Lie algebras}, Phys. Rep. {\bf 71} (1981) 313--400.

\bibitem{KBBT95} I. Krichever, O. Babelon, E. Billey and M. Talon, {\it Spin generalization of the
Calogero-Moser system and the matrix KP equation}, Amer. Math. Soc. Transl. Ser. 2
{\bf 170} (1995) 83--119.

\bibitem{GH84} J. Gibbons and T. Hermsen,
{\it A generalization of the Calogero-Moser system},
Physica D {\bf 11} (1984) 337–-348.

\bibitem{KZ95} I. Krichever and A. Zabrodin, {\it 
Spin generalization of the Ruijsenaars-Schneider model, non-abelian 2D
Toda chain and representations of Sklyanin algebra}, Uspekhi Mat. Nauk
{\bf 50} (1995) 3--56 (in Russian) (English translation: 
Russ. Math. Surv., {\bf 50} (1995) 1101--1150).

\bibitem{RS86} S.N.M. Ruijsenaars and H. Schneider, {\it 
A new class of integrable systems and its relation to
solitons},
 Annals of Physics {\bf 146} (1986) 1--34.
 
\bibitem{Ruij87} S.N.M. Ruijsenaars, {\it Complete integrability of 
relativistic
Calogero-Moser systems and elliptic function identities}, 
Commun. Math. Phys. {\bf 110} (1987) 191--213.

\bibitem{RZ20}
D. Rudneva and A. Zabrodin, {\it Dynamics of poles of 
elliptic solutions to BKP equation},
Journal of Physics A: Math. Theor. {\bf 53} (2020) 075202,
arXiv:1903.00968.

\bibitem{KZ20}
I. Krichever and A. Zabrodin, {\it Kadomtsev-Petviashvili 
turning points and CKP hierarchy},
Commun. Math. Phys. {\bf 386} (2021) 1643--1683,
arXiv:2012.04482.

\bibitem{Z19} A. Zabrodin, {\it Elliptic solutions to integrable nonlinear
equations and many-body systems}, 
Journal of Geometry and Physics {\bf 146} (2019) 103506, arXiv:1905.11383.



\bibitem{K05}
I.M. Krichever, {\it 
Integrable linear equations and the Riemann-Schottky problem}, In: Algebraic geometry and 
number theory. In Honor of Vladimir Drinfeld's 50th birthday. Ed. by Ginzburg, Victor. Basel: 
Birkhäuser. Progress in Mathematics {\bf 253} (2006) 497--514, arXiv:math/0504192.

\bibitem{K10}
I. Krichever, 
{\it Characterizing Jacobians via trisecants of
the Kummer variety}, Annals of Mathematics {\bf 172} (2010)
485--516.

\bibitem{K22}
I. Krichever, 
{\it Abelian pole systems and Riemann-Schottky type systems}, 
arXiv:2202.04585.

\bibitem{KZ22} 
I. Krichever and A. Zabrodin, 
{\it Toda lattice with constraint of type B}, arXiv:2210.12534.

\bibitem{Z21}
A. Zabrodin, {\it How Calogero-Moser particles can stick together},
J. Phys. A: Math. Theor. {\bf 54} (2021) 225201.

\bibitem{KZ21a} I. Krichever and A. Zabrodin, 
{\it Constrained Toda hierarchy and 
turning points of the Ruijsenaars-Schneider model}, 
Letters in Mathematical Physics {\bf 112} 23 (2022), arXiv:2109.05240.

\bibitem{UT84} K. Ueno and K. Takasaki, {\it Toda lattice hierarchy},
Adv. Studies in Pure Math. {\bf 4} (1984) 1--95.

\bibitem{Hirota}
R. Hirota, {\it Discrete analogue of a generalized Toda equation},
J. Phys. Soc. Japan {\bf 50} (1981) 3785--3791.

\bibitem{Miwa}
T. Miwa, {\it On Hirota's difference equations}, Proc. Japan Acad.
{\bf 58} Ser. A (1982) 9--12. 

\bibitem{NRK96} F.W. Nihhoff, O. Ragnisco and V. Kuznetsov,
{\it Integrable time-discretization of the Ruijsenaars-Schneider model},
Commun. Math. Phys. {\bf 176} (1996) 681--700.


\end{thebibliography}
\end{document}